\documentclass[12pt]{article}
\usepackage{epsfig}
\usepackage{amsmath}
\usepackage{hhline}
\usepackage{amssymb}
\usepackage{times}
\usepackage{cite}

\newlength{\dinwidth}
\newlength{\dinmargin}
\begin{document}  
\newcommand{\pom}{{I\!\!P}}
\newcommand{\reg}{{I\!\!R}}
\def\gsim{\,\lower.25ex\hbox{$\scriptstyle\sim$}\kern-1.30ex%
\raise 0.55ex\hbox{$\scriptstyle >$}\,}
\def\lsim{\,\lower.25ex\hbox{$\scriptstyle\sim$}\kern-1.30ex%
\raise 0.55ex\hbox{$\scriptstyle <$}\,}
\newcommand{\PO}{I\!\!P}
\newcommand{\xbj}{x}
\newcommand{\xpom}{x_{\PO}}
\newcommand{\dgr}{^\circ}
\newcommand{\GeV}{\mathrm{GeV}}
\newcommand{\cm}{\mathrm{cm}}
\newcommand{\mm}{\mathrm{mm}}
\renewcommand{\deg}{^\circ}
\newcommand{\qsq}{\ensuremath{Q^2} }
\newcommand{\gevsq}{\ensuremath{\mathrm{GeV}^2} }
\newcommand{\der}{{\mathrm d}}
\begin{titlepage}

\begin{flushleft}
{\tt DESY 09-185    \hfill    ISSN 0418-9833} \\
{\tt October 2009}                  \\
\end{flushleft}
\vspace{36mm}

\begin{center}
\begin{Large}
{\boldmath \bf      
     Measurement of Leading Neutron Production in 
     Deep-Inelastic Scattering at HERA
}

\vspace{2cm}
\rm
H1 Collaboration

\end{Large}
\end{center}

\vspace{2cm}

\begin{abstract}
\noindent
\rm The production of leading neutrons, where the neutron carries a
large fraction $x_L$ of the incoming proton's longitudinal momentum,
is studied in deep-inelastic positron-proton scattering at HERA.
The data were taken with the H1 detector in the years 2006 and 2007 and
correspond to an integrated luminosity of $122~\mathrm{pb}^{-1}$.  The
semi-inclusive cross section is measured 
in the phase space defined by the photon virtuality
$6<Q^2<100~\GeV^2$, Bjorken scaling variable $1.5 \cdot
10^{-4}< x < 3 \cdot 10^{-2}$, longitudinal momentum fraction
$0.32<x_L<0.95$ and neutron transverse momentum \mbox{$p_T<0.2~\GeV$}. 
The leading neutron structure function, $F_2^{LN(3)}(Q^2,x,x_L)$,
and the fraction of deep-inelastic scattering events containing a 
leading neutron are studied as a function of $Q^2$, $x$ and $x_L$.
 Assuming that the pion exchange
mechanism dominates leading neutron production, the data provide
constraints on the shape of the pion structure function.
\end{abstract}
\vspace{5mm}

\it \normalsize 
\begin{center}
Submitted to {\it Eur.Phys.J.C} 
\end{center}

\vspace{1.5cm}

\end{titlepage}

\newpage


\noindent
F.D.~Aaron$^{5,49}$,           
C.~Alexa$^{5}$,                
K.~Alimujiang$^{11,51}$,       
V.~Andreev$^{25}$,             
B.~Antunovic$^{11}$,           
S.~Backovic$^{30}$,            
A.~Baghdasaryan$^{38}$,        
E.~Barrelet$^{29}$,            
W.~Bartel$^{11}$,              
K.~Begzsuren$^{35}$
% ...

\bigskip\noindent{\it
% ...
$ ^{2}$ Vinca  Institute of Nuclear Sciences, Belgrade, Serbia \\
% ...
$ ^{5}$ National Institute for Physics and Nuclear Engineering (NIPNE) ,
          Bucharest, Romania \\
% ...
$ ^{11}$ DESY, Hamburg, Germany \\
% ...
$ ^{25}$ Lebedev Physical Institute, Moscow, Russia$^{ e}$ \\
% ...
$ ^{29}$ LPNHE, Universit\'{e}s Paris VI and VII, CNRS/IN2P3,
          Paris, France \\
 $ ^{30}$ Faculty of Science, University of Montenegro,
          Podgorica, Montenegro$^{ e}$ \\
% ...
$ ^{35}$ Institute of Physics and Technology of the Mongolian
          Academy of Sciences, Ulaanbaatar, Mongolia \\
% ...
$ ^{38}$ Yerevan Physics Institute, Yerevan, Armenia \\
% ...
$ ^{49}$ Also at Faculty of Physics, University of Bucharest,
          Bucharest, Romania \\
% ...
$ ^{51}$ Supported by the Initiative and Networking Fund of the
          Helmholtz Association (HGF) under the contract VH-NG-401. \\

\bigskip
\noindent
% ...
$ ^e$ Supported by the Deutsche Forschungsgemeinschaft \\
% ...
}
\newpage

\section{Introduction}

The production of leading baryons in deep-inelastic scattering (DIS)
provides a testing ground for the theory of
strong interactions in the soft regime.
Events containing a neutron, which carries a large fraction $x_L$
of the longitudinal momentum of the incoming proton, have been 
observed in electron-proton collisions at HERA
\cite{Adloff:1998yg,Chekanov:2002pf,Breitweg:2000nk,Chekanov:2004wn,Aktas:2004gi,Chekanov:2004dk,Chekanov:2007tv}.
The process leading to such events,
$e p \rightarrow e' n X$, is illustrated in Fig.~\ref{diagram1}a.
The pion exchange mechanism, illustrated in  Fig.~\ref{diagram1}b,
is expected to dominate leading neutron production at
large $x_L$ and low transverse momentum of the neutron
\cite{Sullivan:1971kd,Bishari:1972tx,Holtmann:1994rs,Kopeliovich:1996iw,Przybycien:1996zb,Szczurek:1997cw,Khoze:2006hw}.
In this picture of leading neutron production,  the proton fluctuates
into a state consisting of a positively charged pion
and a neutron $p\rightarrow n\pi^+$. The virtual photon subsequently
interacts with a parton from the pion.  
Consequently, the cross section factorises into two parts 
(proton vertex factorisation):
one factor describes the proton fluctuation into a $n\pi^+$ state, 
the other describes the 
photon-pion scattering~\cite{Sullivan:1971kd,Bishari:1972tx}.
The production of leading neutrons in DIS at HERA may therefore provide
constraints on the structure of the pion at low to medium Bjorken-$x$,
while the knowledge of the pion structure from fixed target experiments
\cite{Anassontzis:1987hk}
is limited to higher $x$ values.
Previous H1 and ZEUS studies of semi-inclusive leading neutron DIS cross sections
\cite{Adloff:1998yg,Chekanov:2002pf} demonstrate that these
measurements are indeed sensitive to the structure of the pion and can
distinguish between different parameterisations of the pion structure function.

Energetic neutrons can also be produced in the fragmentation of
the proton remnant. 
The comparison of leading neutron production with inclusive DIS
provides tests of fragmentation mechanisms.
The hypothesis of limiting fragmentation
\cite{Benecke:1969sh,Chou:1994dh}  
states that, in the high-energy limit,
the cross section for the inclusive production of particles in the target fragmentation region 
becomes independent of the incident projectile energy.
This hypothesis implies that, in DIS, leading neutron
production is insensitive to Bjorken-$x$ and the virtuality of the
exchanged photon $Q^2$.  

In this paper a measurement of the semi-inclusive cross section
for leading neutron production in DIS is presented.
This analysis is based on a data sample corresponding to an integrated luminosity which is $36$ times larger than
that of the previous H1 publication~\cite{Adloff:1998yg}.
A new neutron calorimeter with improved performance is used.
The larger data set together with better experimental capabilities
allow the extension of the kinematic range of the
measurement to higher values of $Q^2$ and $x$ and the reduction of the total 
uncertainty of the results.


\section{Event Kinematics and Reconstruction}

\begin{figure}[h]
\epsfig{file=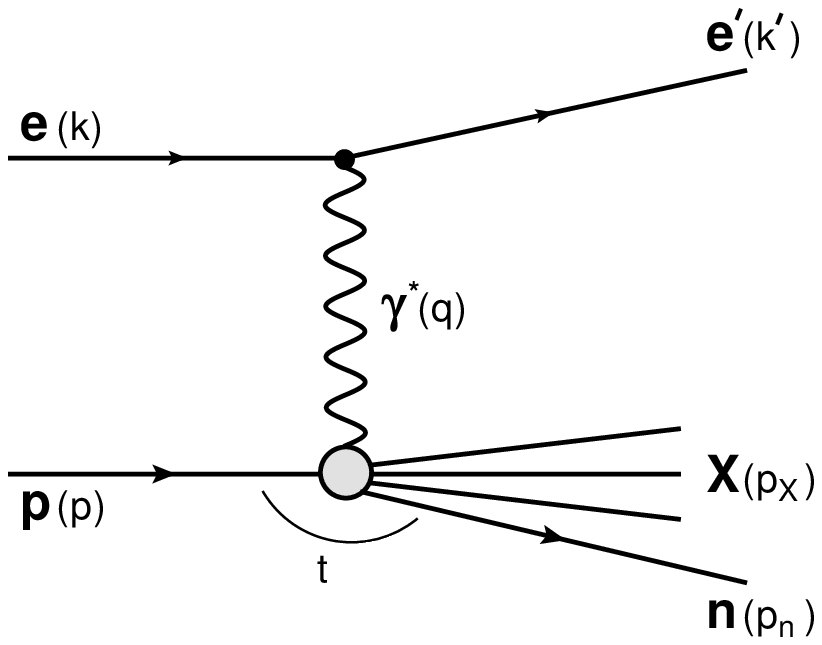,height=58mm}
\hspace*{5mm}
\epsfig{file=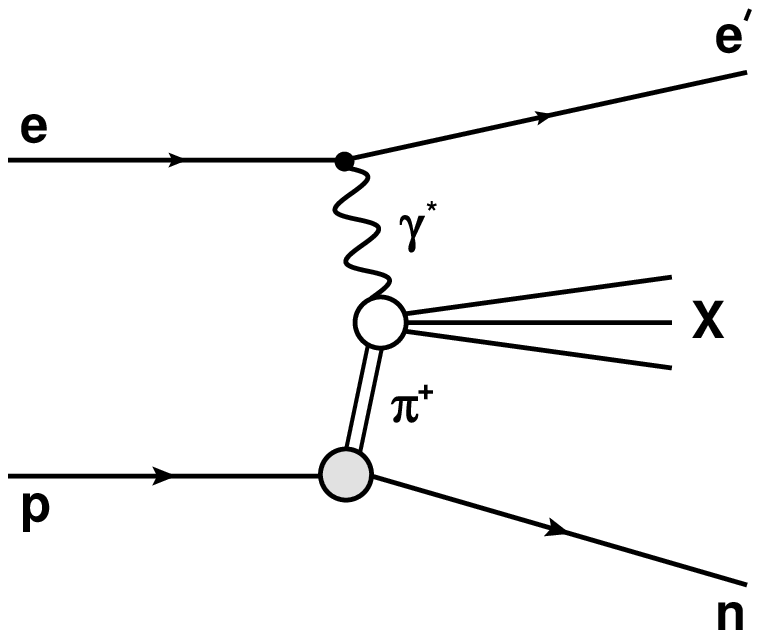,height=58mm}
\caption{ (a) Generic diagram for leading neutron production 
$ep\rightarrow e'nX$ in deep-inelastic scattering.
(b) Diagram of the same process assuming that it proceeds via pion exchange.}

\vspace*{-75mm}
\hspace*{30mm}{\large \bf (a)} \hspace*{74mm} {\large \bf (b)}

\vspace*{73mm}  
\label{diagram1}
\end{figure}

The kinematics of semi-inclusive leading neutron production are shown
in Fig.~\ref{diagram1}a, where the four-vectors of the incoming
and outgoing particles and of the exchanged virtual photon $\gamma^*$
are indicated.
The kinematic variables $Q^2$, $x$ and $y$ are used to describe the inclusive
DIS scattering process. They are defined as
\begin{equation}
Q^2=-q^2\; ,
\hspace{4em}
x=\frac{Q^2}{2 p \cdot q}\; ,
\hspace{4em}
y=\frac{p \cdot q}{p \cdot k}\;,
\end{equation}
where $p$, $k$ and $q$ are the four-momenta of the incident proton,
the incident positron and the exchanged virtual photon, respectively.  
%
These variables are reconstructed using the technique introduced in
\cite{Adloff:1997sc}, which
optimises the resolution throughout the
measured $y$ range by exploiting information from both the scattered
positron and the hadronic final state.

The kinematic variables used to describe the final state neutron 
are the longitudinal momentum fraction $x_L$ and the squared 
four-momentum transfer $t$ 
between the incident proton and the final state neutron
\begin{equation}
   x_L = 1-\frac{q \cdot (p-p_n)}{q \cdot p} \simeq E_n/E_p,
\hspace{2em}  
t = (p-p_n)^2  \simeq  - \frac{p_T^2}{x_L} - (1-x_L)
    {\Bigg (} \frac{m_n^2}{x_L} - m_p^2 {\Bigg )},
\end{equation}
where $m_p$ is the proton mass,
$p_n$ is the four-momentum of the final state neutron,
$m_n$ is the neutron mass and $E_n$ and $p_T$ are
the neutron energy and transverse momentum, respectively.

The four-fold differential cross section for leading neutron production
can be parameterised by a semi-inclusive structure function,
$F_2^{LN(4)}$, defined by
\begin{eqnarray}
  \frac{\der^4 \sigma(e p \rightarrow e n X)}
       {\der x\,\der Q^2\,\der x_L\, \der t} =
   \frac {4 \pi \alpha^2}{x\,Q^4}
   \left ( 1 - y + \frac{y^2}{2} \right )
   F_2^{LN(4)}(Q^2,x,x_L,t).
 \label{eq:eqf2}
\end{eqnarray}

\noindent
The contribution from longitudinally polarised photons can be neglected 
in the phase space studied in this analysis.  
Integrating Eq.~\ref{eq:eqf2} over $t$ yields
the semi-inclusive structure function $F_2^{LN(3)}$ measured in this analysis
\begin{eqnarray}
  \frac{\der^3 \sigma (e p \rightarrow e n X)}
       {\der x\,\der Q^2\,\der x_L} & = & \int_{t_0}^{t_{{{\min}}}}
 \frac{\der^4 \sigma(e p \rightarrow e n X)}
       {\der x\,\der Q^2\,\der x_L\, \der t} \der t \nonumber\\
   & = &
  \frac {4 \pi \alpha^2}{x\,Q^4}
  \left ( 1 - y + \frac{y^2}{2} \right )
   F_2^{LN(3)}(Q^2,x,x_L), 
  \label{eq:eqf23}
\end{eqnarray}
where the integration limits are
\begin{equation}
\label{tlimits}
t_{{\min}}  = - (1-x_L) {\Bigg (} \frac{m_n^2}{x_L} - m_p^2 {\Bigg )}
\hspace{1em}
{\rm and}
\hspace{1em}
t_0 =  -\frac{(p_T^{\max})^2}{x_L}+t_{\min}\,.
\end{equation}
 Here $p_T^{\max}$ is the upper limit of the neutron transverse momentum
 used for the $F_2^{LN(3)}$ measurement.
 Smaller values of $p_T^{\max}$ are expected to enhance the relative contribution of
 pion exchange \cite{Holtmann:1994rs,Kopeliovich:1996iw}. In this analysis $p_T^{\max}$ is 
 set to $0.2~\GeV$, as in the previous H1 publication \cite{Adloff:1998yg}.

\section{Experimental Procedure and Data Analysis}
The data used in this analysis were collected with the
H1 detector at HERA in the years 2006 and 2007 and correspond to an integrated
luminosity of $122~\mathrm{pb}^{-1}$. 
During this period
 HERA collided positrons and protons with energies of $E_e=27.6~\GeV$ and $E_p=920~\GeV$,
respectively.

\subsection{H1 detector}

A detailed description of the H1 detector can be found elsewhere
\cite{Abt:1996hi,Abt:1996xv,Appuhn:1996na,Pitzl:2000wz,Andrieu:1993kh}.  
Here, a brief
account is given of the components most relevant to the present analysis.
The origin of the right-handed H1 coordinate system is the
nominal $ep$ interaction point. The direction of the proton beam
defines the positive $z$-axis (forward direction); the polar angle
$\theta$ is measured with respect to this axis.
Transverse momenta are measured in the $x-y$ plane.

The $ep$ interaction region is surrounded by a two-layered silicon
strip detector and two large concentric drift
chambers.
Using these detectors charged particle momenta are measured  in the
angular range $25\deg<\theta<155\deg$  
with a resolution of 
$\sigma(p_T)/p_T=0.005\, p_T/ \GeV \, \oplus \, 0.015$~\cite{Kleinwort:2006zz}.
The tracking system is surrounded by a finely segmented
Liquid Argon (LAr) calorimeter, which covers a range in polar angle of
\mbox{$4\deg<\theta<154\deg$} with full azimuthal acceptance.  The LAr
calorimeter consists of an electromagnetic section with lead absorber
and a hadronic section with steel absorber.  The total depth of the
LAr calorimeter ranges from $4.5$ to $8$ hadronic interaction lengths.
Its energy resolution, determined in test beam measurements, is
$\sigma(E)/E\approx 12\%/\sqrt{E[\GeV]}\oplus 1\%$ for electrons and
$\sigma(E)/E\approx 50\%/\sqrt{E[\GeV]}\oplus 2\%$ for 
charged pions~\cite{Andrieu:1993tz}.  
The backward region ($153\deg<\theta<177.8\deg$) is covered by a
lead/scintillating-fibre calorimeter, the SpaCal.
Its main purpose is the detection of scattered positrons.  
The energy resolution for positrons is $\sigma(E)/E\approx
7.1\%/\sqrt{E[\GeV]}\oplus 1\%$.
The LAr and SpaCal
calorimeters are surrounded by a superconducting solenoid which
provides a uniform magnetic field of $1.16$~T along the beam direction.

The luminosity is measured via the Bethe-Heitler
process $ep \rightarrow e'p \gamma$. The final state photon is detected in a
dedicated calorimeter situated near the beam pipe at $z=-103~\mathrm{m}$.

\subsection{Detection of leading neutrons}

\begin{figure}[h]
\hspace*{40mm}{\large \bf (a)} \hspace*{70mm} {\large \bf (b)}
\vspace*{-2mm}

   \epsfig{file=d09-185f2a.eps,width=80mm}
   \hspace*{7mm}
   \epsfig{file=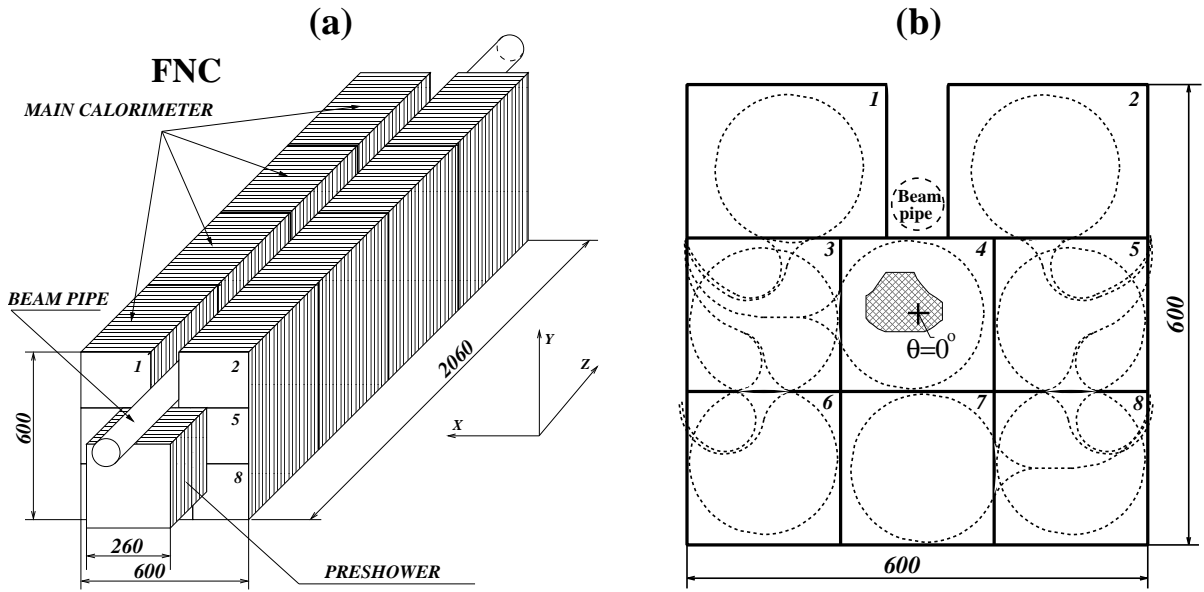,width=70mm}
\vspace*{3mm}

   \caption{(a) Schematic view of the FNC. (b)
             Layout of tiles on an active board of the Main Calorimeter; 
             the position of readout fibres is indicated by dotted lines;
             the hatched area shows the geometrical 
             acceptance window defined by the beam line elements.
             The position corresponding to $\theta=0\deg$ is also indicated. 
             All dimensions are given in $\mm$.}
   \label{gen_view}
 \end{figure}

 Leading neutrons are detected in the forward neutron calorimeter (FNC).
 The FNC is situated at a polar angle of $0\deg$ beyond the magnets
 used to deflect the proton beam, at $z=106~\mathrm{m}$.
 A schematic view of the FNC is shown in Fig.~\ref{gen_view}a.  It consists of the Main
 Calorimeter and the Preshower Calorimeter.
 In addition, two layers of veto counters situated at a distance of
 $2~\mathrm{m}$ in front of the Preshower Calorimeter are used to veto charged particles.

The Preshower Calorimeter is a $40~\cm$ long 
lead-scintillator sandwich calorimeter, corresponding to 
about $60$ radiation lengths or $1.6$ hadronic interaction lengths.
It is composed of $24$ planes: the first $12$ planes  
each consist of a lead plate of $7.5~\mm$ thickness and a scintillator plate of $2.6~\mm$ thickness,
the second $12$ planes
each consist of a lead plate of $14~\mm$ thickness and a scintillator plate of $5.2~\mm$ thickness.
The transverse size of the scintillating plates is \mbox{$26\times26~\cm^2$}.
Each scintillating plate has $45$ parallel grooves holding
$1.2~\mm$ diameter wavelength shifter (WLS) fibres.
In order to obtain good spatial resolution, the orientation of fibres alternates
from horizontal to vertical in consecutive planes. At each plane
the fibres are bundled into nine strips of five fibres. Longitudinally, the strips
are combined into $9$ vertical and $9$ horizontal towers which are finally connected to
$18$ photomultipliers.

The Main Calorimeter of the FNC is
a sandwich-type calorimeter consisting of four identical
sections. Each section is $51.5~\cm$ long 
with transverse dimensions of \mbox{$60\times 60~\cm^2$}
and  consists of $25$ lead absorber plates of $14~\mm$
thickness and $25$ active boards with $3~\mm$ thick scintillators.
Each active board is made of $6$ scintillating tiles with a transverse
size of \mbox{$20\times 20~\cm^2$} and $2$ tiles 
of transverse size \mbox{$20 \times 26~\cm^2$},
as shown in  Fig.~\ref{gen_view}b.
Each  scintillating tile has a circular groove holding a $1~\mm$ diameter WLS fibre
which is attached via light connectors to two $1~\mm$ diameter transparent fibres.
Longitudinally, for each section
the fibres from $25$ tiles with the same transverse position
are bundled into a tower and connected to one photomultiplier.
There are then $8$ towers in each section, making a total of
$32$ towers in the Main Calorimeter.
The proton beam pipe is located in a rectangular space along the top of the
calorimeter, as can be seen in Fig.~\ref{gen_view}.
The total length of the Main Calorimeter is $206~\cm$, corresponding to 
$8.9$ hadronic interaction lengths.

All modules of the FNC were initially calibrated at CERN using
electron beams with energies between $120$ and $230~\GeV$ and
hadron beams with energies between $120$ and $350~\GeV$.
The FNC was positioned on a movable
platform which allowed the response 
of each module and tower to be measured separately.
After this initial calibration, the FNC had an approximately uniform
response independent of impact position.
The linearity of the energy response of the FNC was measured at HERA
from beam-gas interactions
during dedicated runs, 
when the proton beam was
accelerated to five intermediate energies between $150~\GeV$ and $920~\GeV$.
The energy response of the FNC is linear to a precision of $3\%$
and the hadronic energy resolution is
$\sigma(E)/E\approx 63\%/\sqrt{E[\GeV]}\oplus 3\%$.
The energy resolution for electromagnetic showers, which are always fully
contained in the Preshower Calorimeter,  
is $\sigma(E)/E \approx 20\%/\sqrt{E~[\GeV]}$.
For hadronic showers starting in the Main Calorimeter,
the spatial resolution is
$\sigma(x,y) \approx 10~\cm/\sqrt{E~[\GeV]}\oplus 0.6~\cm$.
A better spatial resolution of about $2~\mm$ is achieved for the 
electromagnetic showers and for those hadronic showers 
starting in the Preshower Calorimeter.

The acceptance of the FNC is defined by the aperture of the
HERA beam line magnets and is limited to neutron scattering
angles of $\theta_n\lsim 0.8~\mathrm{mrad}$ with an approximately
$30\%$ azimuthal coverage.
The geometrical acceptance window of the FNC is indicated
in Fig.~\ref{gen_view}b.

After the calorimeter was installed in the H1 beam line, 
the stability of calibration constants 
was monitored  using interactions between the proton beam and 
residual gas molecules in the beam pipe.
The neutron energy spectrum was compared with
the results of Monte Carlo simulation based on pion exchange.
From this monitoring, the time dependent variations of the calibration constants of order of
a few percent were determined.
%
%
%
Short term variations of photomultiplier gain were monitored using
LED signals.  The LEDs were operated
during empty bunches at a frequency of $0.8~\mathrm{Hz}$.  The averaged LED signal
responses were used to provide offline energy corrections applied
during the reconstruction.

The longitudinal segmentation of the FNC allows efficient discrimination
of neutrons from charged particles or electromagnetic energy deposits.
Charged particles are rejected using signals from the veto counters.
Events with an energy deposit only in the Preshower Calorimeter are 
identified as electromagnetic showers (photons, $\pi^0$ mesons). 
All other patterns of energy deposition are identified as hadronic showers, which are
subsequently classified as either those starting in the Preshower Calorimeter 
or those starting in the Main Calorimeter, the latter having worse spatial
resolution. In this analysis both hadronic shower types are used.

\subsection{Event selection }

The data
sample was collected using a trigger which requires the scattered
positron to be measured in the SpaCal and 
at least $100~\GeV$ energy to be deposited in the FNC.  
The trigger signal from the FNC is formed using an analogue sum of signals 
from the Preshower Calorimeter and the central towers of the Main Calorimeter.
The trigger efficiency is above $97\%$ in most
of the analysis phase space, decreasing to $94\%$ for $x_L<0.4$. 

The selection of DIS events is based on the identification
of the scattered positron as the most energetic compact calorimetric deposit 
in the SpaCal  with an
energy $E_e'>11~\GeV$ and a polar angle $156\deg<\theta_e'<175\deg$.
The energy weighted cluster radius is required to be less than $4~\cm$, 
as expected for an electromagnetic shower~\cite{Aaron:2009kv}.
The $z$-coordinate of the primary event vertex is required to be 
within $\pm 35~\cm$ of the nominal position of the interaction point. 
The remaining clusters in the calorimeters and the charged tracks are 
combined to reconstruct the hadronic final state.
%
To suppress events with initial state hard photon radiation, as well as
events originating  from  non-$ep$ interactions,  
the quantity $E - p_z$, summed over  
all reconstructed particles including the positron, is required to lie 
between  $35~\GeV$ and $70~\GeV$.
This quantity is expected to be twice the electron beam energy
for DIS events without QED radiation.
Furthermore, events are
selected within the kinematic range $6<Q^2<100~\GeV^2$, $0.02<y<0.6$
and $1.5 \cdot 10^{-4}<x<3\cdot 10^{-2}$.

Events containing a leading neutron are selected by requiring a
hadronic cluster in the FNC with an energy above $275~\GeV$ and a
polar angle below $0.75~\mathrm{mrad}$. 
The cut on polar angle, defined by the geometrical acceptance of the FNC, 
restricts the neutron transverse momenta $p_T$ to the range
$p_T< x_L \cdot 0.69~\GeV$.

The final data sample contains $315960$ events which satisfy these selection criteria.
For the measurement of $F_2^{LN(3)}$, an additional requirement on the 
transverse momentum of the neutron $p_T<0.2~\GeV$ 
is applied to enhance the relative contribution of pion exchange
and to avoid having an $x_L$ dependent $p_T$ cut.
The number of events selected with this additional requirement is $209150$.

\subsection{Monte Carlo simulation and corrections to the data}
\label{sec:mc}

Monte Carlo simulations are used to correct the data for the effects
of detector acceptance, inefficiencies and migrations between measurement
intervals due to finite resolution and QED radiation.  All generated
events are passed through a GEANT3~\cite{Brun:1978fy} based simulation
of the H1 apparatus and are processed using the same reconstruction 
and analysis framework as is used for the data.

The DJANGO~\cite{Charchula:1994kf}
program generates inclusive DIS events.
It is based on leading order electroweak cross sections and takes QCD
effects into account up to order $\alpha_s$.  The hadronic
final state is simulated using ARIADNE \cite{Lonnblad:1992tz}, based on the
Colour Dipole Model, with subsequent hadronisation effects modelled
using the Lund string fragmentation model as implemented in JETSET~\cite{JETSET}.
DJANGO is also used in this analysis to simulate events where leading neutrons
originate from proton remnant fragmentation.
RAPGAP~\cite{Jung:1993gf} is a general purpose event generator for inclusive and
diffractive $ep$ interactions.  Higher order QCD effects are simulated
using parton showers and the final state hadrons are
obtained via Lund string fragmentation.  Higher order electroweak
processes in the DJANGO and RAPGAP generators are simulated using an
interface to HERACLES~\cite{Kwiatkowski:1990es}.

In the version denoted below as RAPGAP-$\pi$, the program
simulates exclusively the scattering of virtual or real photons
off an exchanged pion. 
Here, the cross section for photon-proton scattering to the final state 
$nX$ takes the form
\begin{equation}
  \der\sigma (e p\rightarrow e'nX) =f_{\pi^+/p}(x_L,t)\cdot
  \der\sigma(e\pi^+\rightarrow e'X),
\label{crsec}
\end{equation}
where $f_{\pi^+/p}(x_L,t)$ represents the pion flux associated with the 
splitting of a proton into a $\pi^+ n$ system
and $\der\sigma(e\pi^+\rightarrow e'X)$ is the cross section of
the positron-pion  interaction.
 There are several parameterisations of the pion flux
 \cite{Bishari:1972tx,Holtmann:1994rs,Kopeliovich:1996iw,Przybycien:1996zb,Szczurek:1997cw}.
 In this analysis, the pion flux factor is taken from the light-cone  representation 
 \cite{Holtmann:1994rs} as
\begin{equation}
 f_{\pi^+/p}(x_L,t)=\frac{1}{2\pi}\frac{g^2_{p\pi n}}{4\pi}
(1-x_L)\frac{-t}{(m_\pi^2-t)^2}
  \exp \left(- R^2_{\pi n} \frac{m^2_\pi-t}{1-x_L}\right),
\label{holtmanflux}
\end{equation}
where $m_\pi$ is the pion mass,
$g^2_{p\pi n}/4\pi=13.6$ is the $p\pi n$ coupling constant
 deduced from phenomenological analyses of low-energy data~\cite{Timmermans:1990tz} 
 and  $R_{\pi n}=0.93~\GeV^{-1}$
is the radius of the pion-proton Fock state~\cite{Holtmann:1994rs}.

The DJANGO and RAPGAP-$\pi$ Monte Carlo simulations are calculated using 
GRV leading 
order parton distributions for the proton \cite{Gluck:1991ng} and the pion 
\cite{Gluck:1991ey}, respectively.

\begin{figure}[p]
\epsfig{file=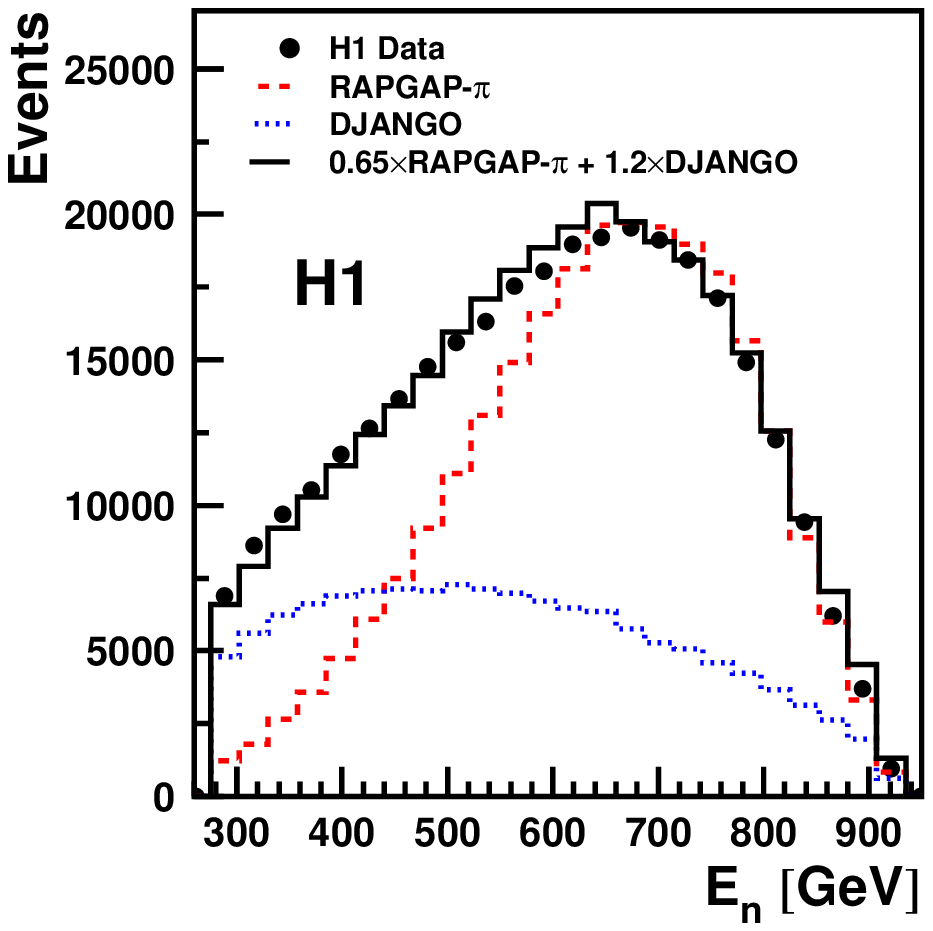,width=80mm} 
\epsfig{file=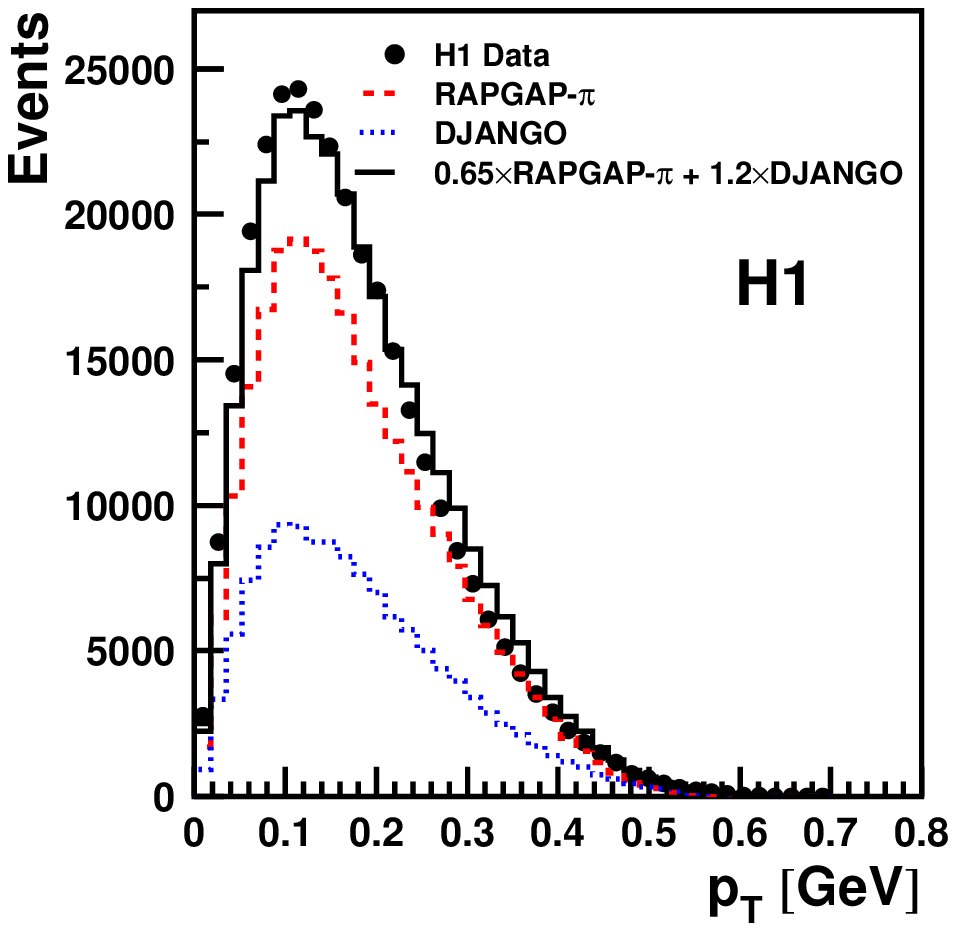,width=80mm} 
\begin{picture}(1,1)
\put(67.,76){\large \bf (a)}
\put(148.,76){\large \bf (b)}
\end{picture}
\caption{The observed neutron energy (a) and transverse momentum (b)
distributions in the kinematic range
 $6< Q^2 < 100~\GeV^2$ and $1.5\cdot 10^{-4} < x < 3\cdot 10^{-2}$.
The data are compared to the predictions of RAPGAP-$\pi$  (dashed line) 
and DJANGO (dotted line) Monte Carlo simulations.  
Also shown is a weighted combination of those two simulations (full line), as
described in Sect.~\ref{sec:mc}.
}
\label{fig:En}
\end{figure}

Figure~\ref{fig:En} shows the observed energy and $p_T$ distribution
of the neutron for selected data sample together with Monte Carlo
simulations.
The differences between the
RAPGAP-$\pi$ and DJANGO generators are particularly 
visible in the neutron
energy distributions (Fig.~\ref{fig:En}a). The RAPGAP-$\pi$ simulation
peaks at $E_n \sim 650~\GeV$ and describes the shape of the
distribution at high energies.  
At lower energies ($E_n \lsim 600~\GeV$) the RAPGAP-$\pi$ simulation 
does not describe the data.
In this region additional 
physics processes which are expected to contribute significantly 
are not simulated.
In contrast, the DJANGO simulation predicts a large contribution 
to the cross section in this region. The best description of the data is achieved if
the predictions of the RAPGAP-$\pi$ and DJANGO Monte Carlo programs
are added, using weighting factors
of $0.65$ and $1.2$ for RAPGAP-$\pi$ and DJANGO, respectively.
This Monte Carlo combination is labelled as 
``$0.65\times$RAPGAP-$\pi~+~1.2\times$DJANGO''  in the figures
and is used to correct the data.

Cross sections at hadron level are determined from selected events 
by applying bin dependent correction factors. These factors
are determined 
from  the combination of DJANGO and RAPGAP-$\pi$
Monte Carlo simulations as the ratios of the cross sections 
obtained from particles at hadron level without QED radiation
to the cross section calculated using reconstructed particles
and including QED radiation effects.
The typical value of these factors is about $1.2$ at the highest $x_L$
increasing to $4$ at the lowest $x_L$ due to the non-uniform azimuthal
acceptance of the FNC.

The binning in $Q^2$, $x$ and $x_L$ used to measure $F_2^{LN(3)}$ is
given in Table~\ref{tab:binlim}.  
The bin purities,  defined as the fraction of events reconstructed
in a particular bin that originated from that bin on hadron level, 
and the bin stabilities,  defined as the fraction of events originating
from a particular bin on hadron level that are reconstructed in that bin,
are higher than $50\%$ for $(Q^2,x)$-bins and higher
than $60\%$ for $x_L$ bins. 

The measured distributions may contain background arising from
different sources.
The background from photoproduction processes, where the positron is
scattered into the backward beam pipe and a particle from the hadronic
final state fakes the positron signature in the SpaCal, is estimated
using the PHOJET Monte Carlo generator~\cite{PHOJET}. This background is negligible
except at the highest $y$ values where it can reach $6\%$.  
Background also arises from the random coincidence of DIS events, causing
activity in the central detector, with proton beam-gas interactions,
which give a neutron signal in the FNC.  This contribution is
estimated by combining DIS events with neutrons
originating from beam-gas interactions in the bunch-crossings adjacent
to the bunch-crossing of the DIS event.  It is found to be smaller than $1\%$.
The estimated background contributions are not subtracted from the measurements.

The contribution from proton dissociation, where the leading neutron
originates from the decay of a higher mass state, is estimated using
an implementation in RAPGAP of the dissociation model originally
developed for the DIFFVM \cite{List:1998jz} Monte Carlo generator. 
This contribution can be up to $30\%$ at low $x_L$ values, but is less 
than $5\%$ for $x_L>0.7$.
It is included in the cross section definition.

\subsection{Systematic uncertainties}

The effects of various systematic
uncertainties on the cross section measurements are determined using
Monte Carlo simulations, by propagating the corresponding estimated
measurement uncertainty through the full analysis chain.

The acceptance of the FNC is defined by the interaction
point and the geometry of the HERA magnets and is determined using
Monte Carlo simulations.  The angular distribution of the neutrons
studied in this analysis is sharply peaked in the forward direction.
Therefore the acceptance depends critically on small
inclinations of the incoming proton beam with respect to its nominal
direction. The uncertainty on the position of the neutron impact point 
is estimated to be $5~\mm$, which results in a
$4.4\%$ uncertainty in the FNC acceptance.  The
uncertainty in the neutron detection efficiency and the uncertainty of
$2\%$ on the absolute energy scale of the FNC lead to a 
systematic error on the cross section of  $2\%$ and $6\%$, respectively.
An additional $0.5\%$ uncertainty is attributed to the trigger efficiency.
These effects are strongly correlated between measurement intervals and
mainly contribute to the overall normalisation uncertainty.

The uncertainties on the measurements of the scattered positron energy
($1\%$) and angle ($1~\mathrm{mrad}$) in the SpaCal lead to a 
combined systematic uncertainty of typically $1.9\%$ on the cross section.  
The uncertainty of the energy measurement of the hadronic final 
state in the central detectors affect the reconstruction 
of the kinematic variables $y$, $Q^2$ and $x$. This hadronic energy scale
uncertainty is
estimated to be $4\%$ for this measurement, leading to an
 uncertainty on the cross section of $0.4\%$ on average.

The systematic error on the efficiency to reconstruct the event vertex
is determined by comparing the reconstruction efficiencies for the data 
and the Monte Carlo simulation.  The discrepancy is less than $1\%$.

The systematic uncertainty arising from the radiative corrections 
and the model dependence of the data correction are estimated by varying
the DJANGO and RAPGAP-$\pi$ scaling factors described in
Sect.~\ref{sec:mc} within values permitted by the data.  The
resulting uncertainty on the cross section is below $5\%$ in most 
of the bins and typically $2\%$.

The luminosity measurement uncertainty for the selected run
period leads to an overall normalisation uncertainty of $5\%$.

The total systematic error in each bin is calculated
as the quadratic sum of all contributions. 
Systematic errors are typically $10\%$ for $F_2^{LN(3)}$ and
$14\%$ for $\der\sigma/\der x_L$ measurements. 

\section{Results}
\label{sec:result}

\begin{figure}[p]  
\epsfig{file=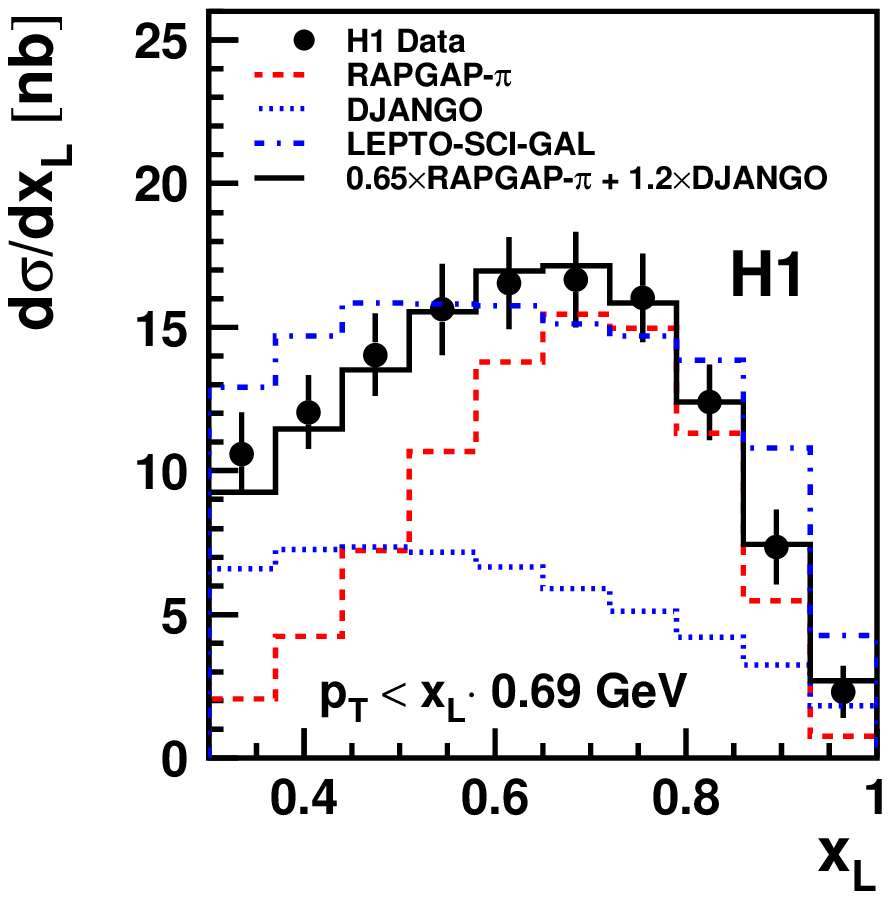,width=80mm}
\epsfig{file=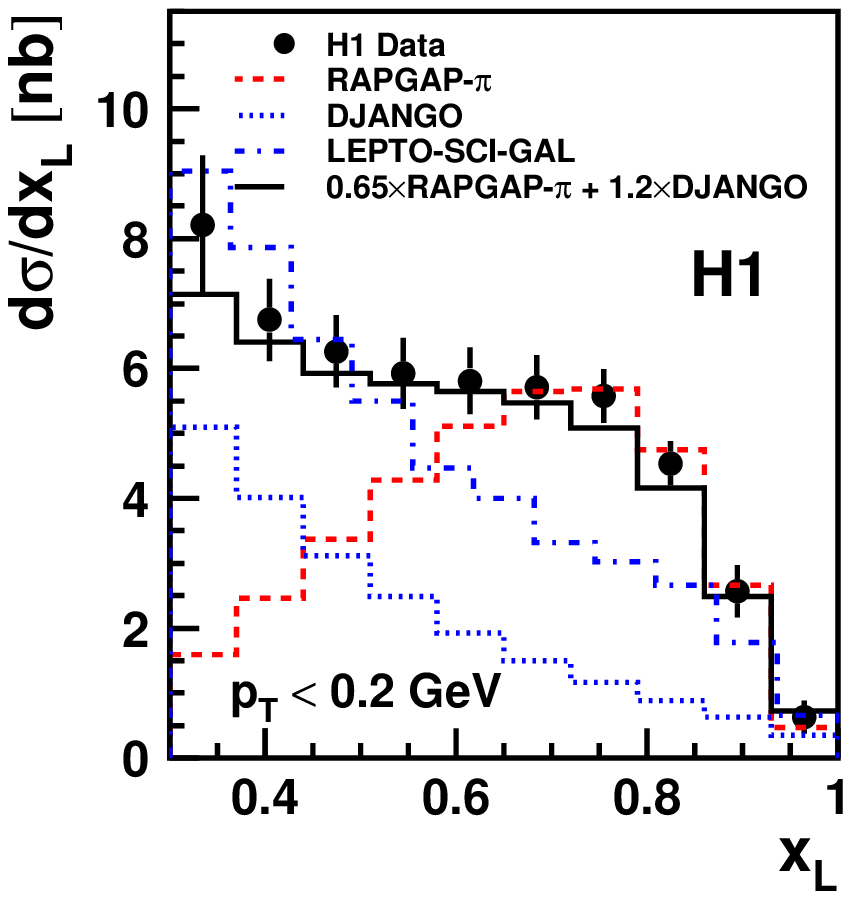,width=80mm}
\begin{picture}(1,1)
\put(67.,75){\large \bf (a)}
\put(147.,75){\large \bf (b)}
\end{picture}
\caption{The cross section as a function of the fractional energy 
of the neutron $x_L$  in the kinematic range
 $6< Q^2 < 100~\GeV^2$ and $1.5\cdot 10^{-4} < x < 3\cdot 10^{-2}$. 
The transverse momentum of the neutron
is restricted to $p_T<x_L \cdot 0.69~\GeV$ (a) and $p_T < 0.2~\GeV$ (b).
The data are compared to the predictions of RAPGAP-$\pi$  (dashed line), 
DJANGO (dotted line) and LEPTO-SCI-GAL (dash-dotted line) Monte Carlo simulations.  
Also shown is a weighted combination of RAPGAP-$\pi$ and DJANGO 
simulations (full line), as described in Sect.~\ref{sec:mc}.   
}
\label{fig:EnCrsec}
\end{figure}

The single differential leading neutron DIS cross sections as a
function of $x_L$ are presented in Fig.~\ref{fig:EnCrsec} and Table~\ref{tab:dsdxl1}.
In Fig.~\ref{fig:EnCrsec}b the differential cross section in $x_L$ is shown
for $p_T<0.2~\GeV$.  
The difference between the shapes of distributions in Fig.~\ref{fig:EnCrsec}a
and Fig.~\ref{fig:EnCrsec}b, in particular a steep fall 
at low $x_L$ in Fig.~\ref{fig:EnCrsec}a,  
is due to the geometrical acceptance of the FNC 
which restricts the accessible $p_T$ range to $p_T<x_L\cdot 0.69~\GeV$.

The measured cross sections are compared with the Monte Carlo simulations.
For large values of $x_L \gsim 0.7$, 
the RAPGAP-$\pi$ simulation describes the shape of the $x_L$
distributions well, in agreement with the assumption that at high $x_L$
the dominant mechanism for leading neutron production is pion
exchange.
The full $x_L$ range is well described 
by the sum of the RAPGAP-$\pi$ and DJANGO Monte Carlo generators, using
the scaling factors discussed in Sect.~\ref{sec:mc}.
This indicates that the $\pi$ exchange mechanism dominates at high $x_L$, while proton remnant fragmentation gives a significant contribution at low $x_L$.

The measured cross sections are also compared with the predictions
of the Soft Colour Interaction model (SCI) \cite{Edin:1995gi},
implemented in the LEPTO Monte Carlo generator program  \cite{Ingelman:1996mq}.
In the SCI model, the production of leading baryons and diffraction-like configurations 
is enhanced via non-perturbative colour rearrangements between the outgoing partons.
A refined version of the model uses a generalised area law (GAL) \cite{Rathsman:1998tp}
for the colour rearrangement probability.
Compared to DJANGO MC predictions, SCI-GAL
improves the description at higher $x_L$,
as can be seen in Fig.~\ref{fig:EnCrsec}.
 However, for lower $p_T$ values 
the predicted cross section for $x_L>0.5$ is still too low.

\begin{figure}[p]
\hspace*{-10mm}
\epsfig{file=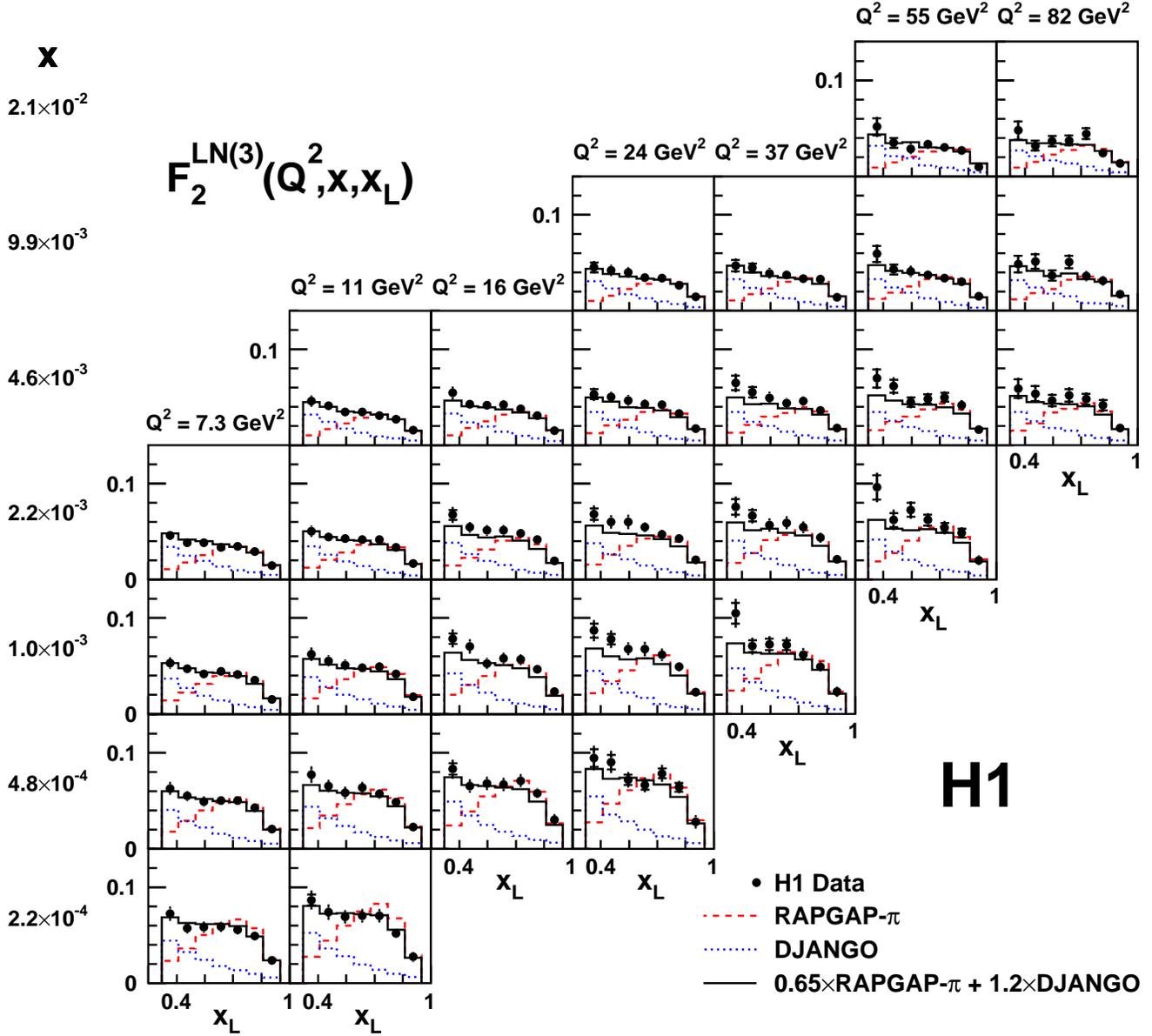,width=180mm}
\caption{The semi-inclusive structure function $F_2^{LN(3)}(Q^2,x,x_L)$, for
neutrons with \mbox{$p_T < 0.2~\GeV$},
compared to the predictions of RAPGAP-$\pi$ (dashed line) and 
DJANGO (dotted line) Monte Carlo simulations.  Also shown is a weighted 
combination of those two simulations (full line), as
described in Sect.~\ref{sec:mc}. 
}
\label{FNC-F2-MC}
\end{figure}

Figure~\ref{FNC-F2-MC} and Tables~\ref{tab:f2ln3p1}-\ref{tab:f2ln3p4}
present the measurement of the semi-inclusive
structure function $F_2^{LN(3)}(Q^2,x,x_L)$ in the range
$6<Q^2<100~\GeV^2$, $1.5\cdot 10^{-4}<x<3\cdot 10^{-2}$,
$0.32<x_L<0.95$ and $p_T<0.2~\GeV$.  
In all $(Q^2,x)$ bins, the shape of the $F_2^{LN(3)}$
distribution as a function of $x_L$ is similar to the shape of the 
single differential
cross section in $x_L$ (Fig.~\ref{fig:EnCrsec}b).  The distributions
are reasonably well described by the combination of RAPGAP-$\pi$ and DJANGO
simulations.

\begin{figure}[p]
\hspace*{-10mm}
\epsfig{file=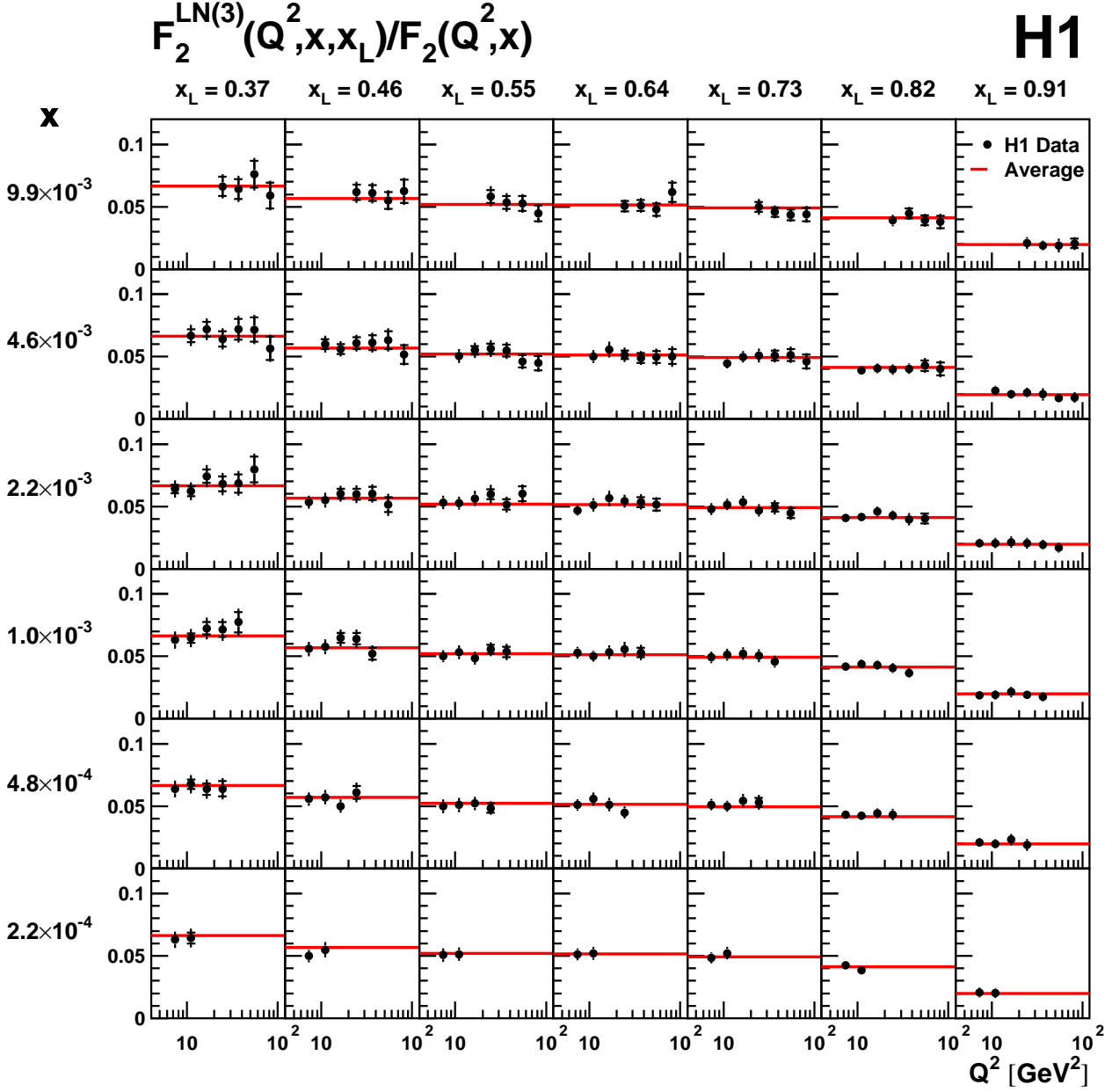,width=170mm}
\caption{The ratio of the semi-inclusive structure function
$F_2^{LN(3)}(Q^2,x,x_L)$, for
neutrons with \mbox{$p_T < 0.2~\GeV$}, to the proton structure function 
$F_2(Q^2,x)$ obtained from the H1PDF2009 fit to inclusive DIS data~\cite{Aaron:2009kv}.
The lines show the average value for a given $x_L$ bin.}
\label{FNC-ratio}
\end{figure}

The measurement of $F_2^{LN(3)}$ allows the validity of the hypothesis
of limiting fragmentation to be tested, according to which the production of leading
neutrons in the proton fragmentation region
is independent of $Q^2$ and $x$.
To investigate this prediction, the ratio of the semi-inclusive structure
function $F_2^{LN(3)}$ to the proton structure function $F_2$
is studied  as a function of $Q^2$ in bins of $x$ and $x_L$
(Fig.~\ref{FNC-ratio}).
The values of $F_2$ are obtained from the
H1PDF2009 parameterisation~\cite{Aaron:2009kv}.  
The horizontal lines indicate the average value of the ratio for 
a given $x_L$ bin. 
These average values decrease from $7\%$  to $2\%$
with increasing $x_L$, reflecting the general behaviour observed 
in Figs.~\ref{fig:EnCrsec} and \ref{FNC-F2-MC}.
The ratios are almost independent of $x$ and $Q^2$ in each $x_L$ bin,
implying that $F_2^{LN(3)}$ and $F_2$ have a similar
$(Q^2,x)$ behaviour, as expected from the hypothesis of limiting
fragmentation.  

\begin{figure}[p]
\hspace*{-10mm}
\epsfig{file=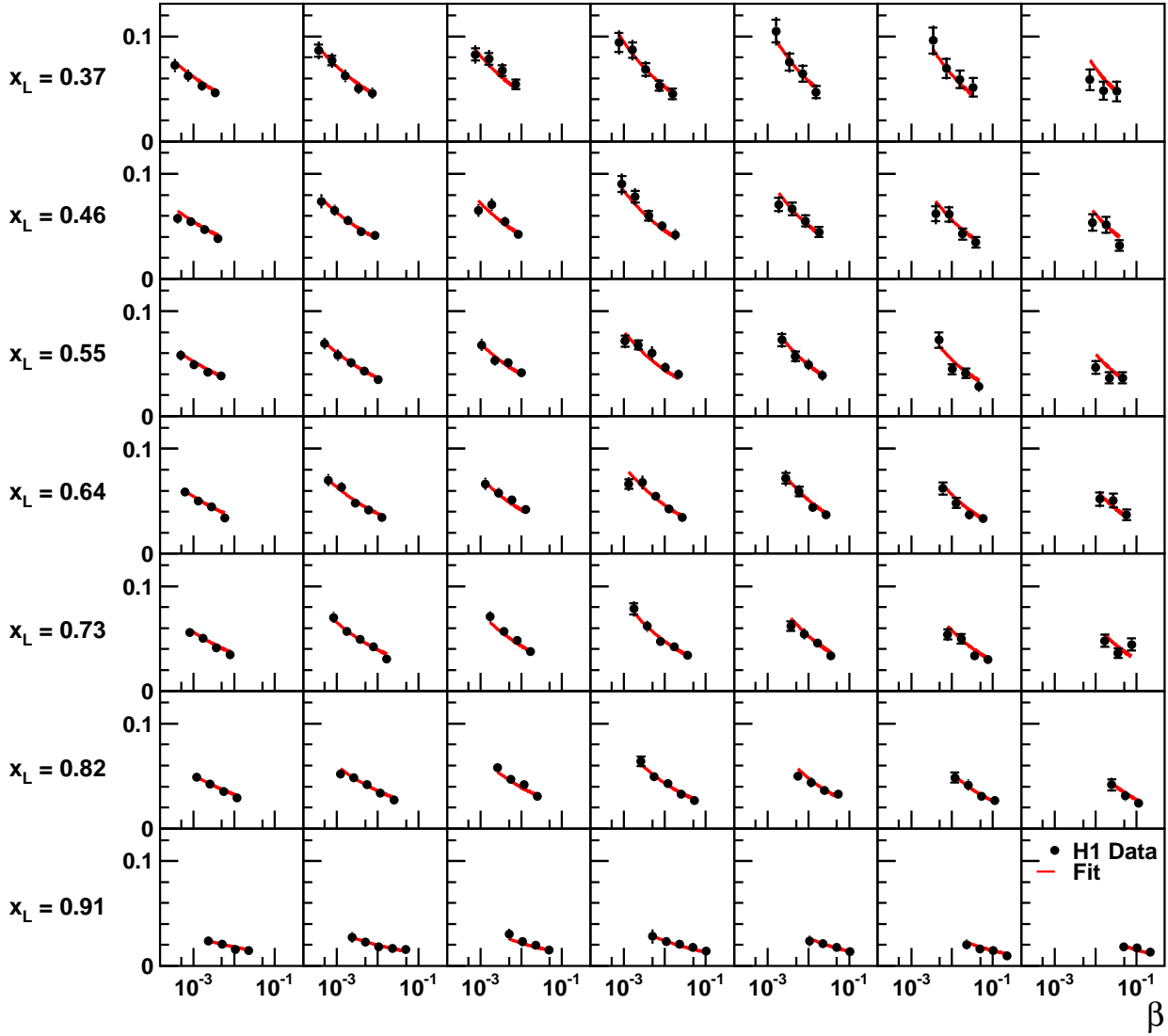,width=170mm}   
\caption{The semi-inclusive structure function $F_2^{LN(3)}(Q^2,\beta,x_L)$, for
neutrons with \mbox{$p_T < 0.2~\GeV$},
shown as a function of $\beta$ in bins of $Q^2$ and $x_L$.
The lines are the results of the fit with a 
function $c(x_L)\cdot \beta^{-\lambda(Q^2)}$ as described in Sect.~\ref{sec:result}.
}
\label{FNC-F2VSBETA}
\end{figure}

Assuming that leading neutrons are produced
via the exchange of a colour singlet particle, e.g. the $\pi^+$,
the structure function $F_2^{LN(3)}$
factorises into a flux factor which is a function of $x_L$ and
a structure function  $F_2^{LN(2)}$ which depends on $Q^2$ and $\beta=x/(1-x_L)$. 
The quantity $\beta$ can be interpreted as
the fraction of the exchanged particle's momentum carried by the  parton interacting
with the virtual photon.
The value of $F_2^{LN(3)}(Q^2,\beta,x_L)$
is obtained by replacing the variable $x$ with $\beta$ in Eq.~\ref{eq:eqf23}. 
The distribution of $F_2^{LN(3)}$, shown in  Fig.~\ref{FNC-F2VSBETA}, 
has a similar dependence on $\beta$  in all $(Q^2,x_L)$ bins.
This behaviour 
can be approximated by a power law function $F_2^{LN(3)} \propto \beta^{-\lambda}$.
In each $(Q^2,x_L)$ bin a fit of the parameter $\lambda$ is performed.
Within uncertainties the value of the fitted parameter
$\lambda$ is independent of $x_L$ 
which is consistent with proton vertex factorisation.

However, as a function of $Q^2$, the values of $\lambda$ increase
from $0.23$ at lowest $Q^2$ to $0.3$ at highest $Q^2$.
This slow $Q^2$ dependence is similar to the rise
towards low $x$ of the proton structure function $F_2$
 measured in inclusive DIS \cite{Adloff:2001rw}.
It is further investigated by fitting the measured $F_2^{LN(3)}(Q^2,\beta,x_L)$ 
assuming the following functional form
\begin{equation}
F_2^{LN(3)}(Q^2,\beta,x_L)=c(x_L) \cdot \beta^{-\lambda(Q^2)}
\hspace*{5mm} {\rm with} \hspace*{5mm} \lambda=a\cdot \ln(Q^2/\Lambda^2)
\label{eq:fit1}
\end{equation}
where $a$, $\Lambda$ and the normalisations $c(x_L)$
are the free parameters of the fit. 
This nine parameter fit over the whole $x_L$ range has a 
good $\chi^2$, supporting the validity of the employed ansatz 
(Eq.~\ref{eq:fit1}). Within the experimental uncertainties the 
obtained values of the parameters $a = 0.052 \pm 0.003$ and 
$\Lambda = 0.416 \pm 0.052~\GeV$ are in reasonable agreement 
with those obtained in the analysis of the proton 
structure function $F_2$~\cite{Adloff:2001rw}. This demonstrates the 
similarity between the $Q^2$ evolution of $F_2^{LN(3)}$ and the 
$Q^2$ evolution of the proton structure function $F_2$.

Since pion exchange dominates leading neutron production at high $x_L$
and low $p_T$, the measurement of $F_2^{LN(3)}$ in the range $0.68<x_L<0.77$ 
can be used to estimate the pion structure function at low  Bjorken-$x$,
following the procedure introduced in \cite{Adloff:1998yg}.
Assuming proton vertex factorisation, which is supported by the present
data as explained above, 
the quantity $F_2^{LN(3)}/\Gamma_\pi$ can be
interpreted as being equal to the structure function of the pion, where
\begin{equation}
   \Gamma_{\pi}(x_L) = \int_{t_0}^{t_{\min}}
      f_{\pi/p}(x_L,t) \, \der t 
\end{equation}
is the integral of the pion flux over the measured $t$-range, where
$t_0$ and $t_{\min}$ are given by Eqs.~\ref{tlimits}.
The pion flux from Eq.~\ref{holtmanflux} used for the RAPGAP-$\pi$
simulation yields $\Gamma_{\pi}=0.13$ for $p_T^{\max}=0.2~\GeV$
at $x_L=0.73$, which is the central value of the chosen $x_L$ range.  
Using other parameterisations 
of the pion flux, e.g. from 
\cite{Bishari:1972tx,Kopeliovich:1996iw,Przybycien:1996zb,Szczurek:1997cw},
leads to values of the pion flux integral which may
differ by up to $30\%$.
In this evaluation of the pion structure function contributions from background processes like 
the exchange of $\rho$ and $a_2$-mesons, proton diffractive dissociation and 
$\Delta$ production are not taken into account. 
Within the narrow $x_L$ range considered here they are only expected to affect the 
absolute normalisation of the results.
The contribution of neutrons from fragmentation is of order $25-35\%$,
as can be estimated using the DJANGO prediction.
The relative size of this contribution is largely independent of 
$Q^2$ and $\beta$ and 
thus has little impact on the shape of the distribution.

\begin{figure}[p]
\centering
\epsfig{file=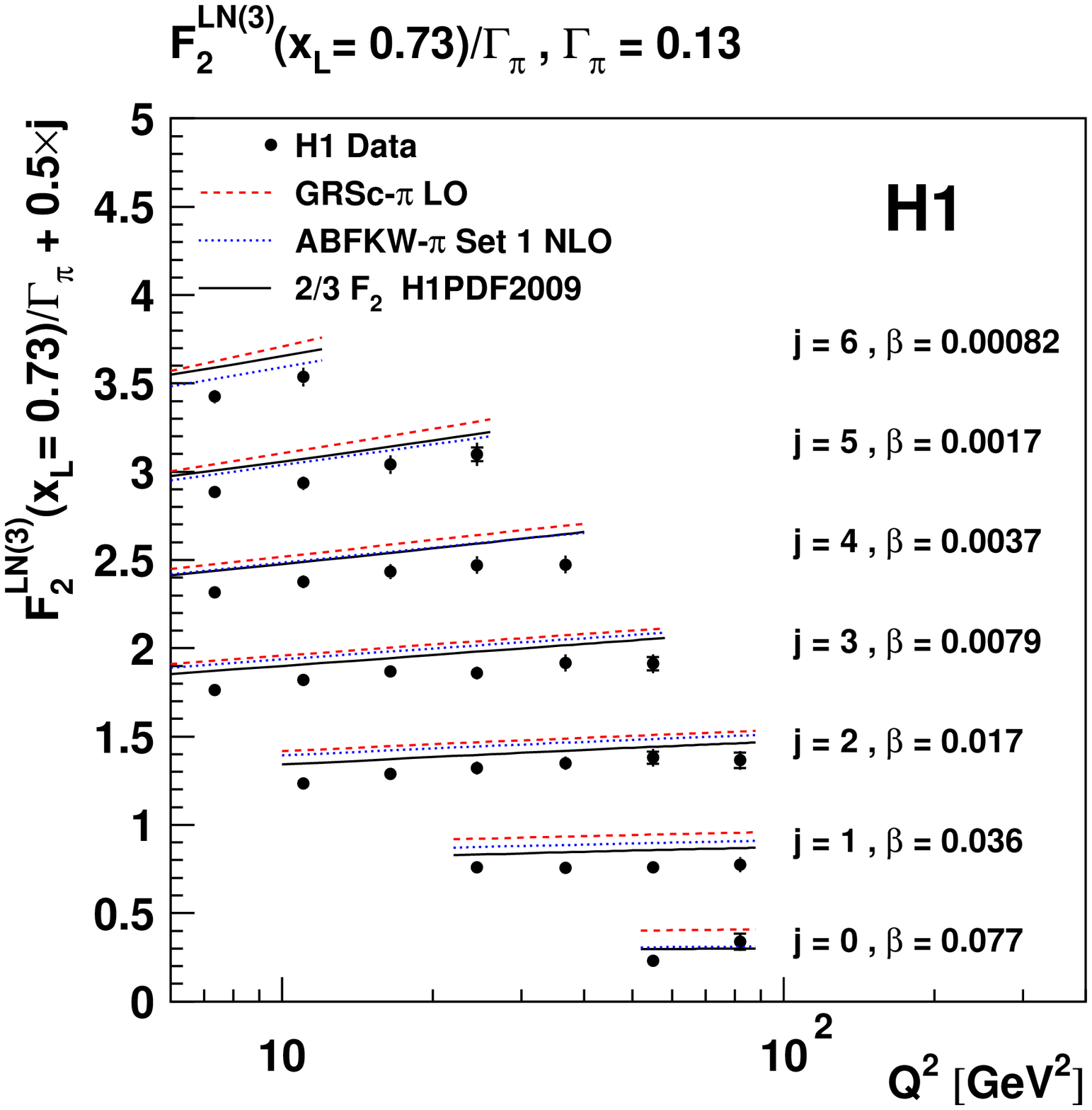,width=140mm}
\caption{
The semi-inclusive structure function $F_2^{LN(3)}$, for
neutrons with $p_T < 0.2~\GeV$,
divided by the pion flux $\Gamma_\pi$ integrated over
$t$ at the central value $x_L=0.73$, shown
as a function of $Q^2$ in bins of $\beta$.
The pion flux is defined in Eq.~\ref{holtmanflux}. 
The data are compared to two different parameterisations of the
pion structure function $F_2^{\pi}$~\cite{Gluck:1999xe,Aurenche:1989sx} 
and to the H1PDF2009 parameterisation of 
the proton structure function \cite{Aaron:2009kv}, which has been scaled by 2/3.
}
\label{FNC-F2PIQ2}
\end{figure}

\begin{figure}[p]
\hspace*{-10mm}
\epsfig{file=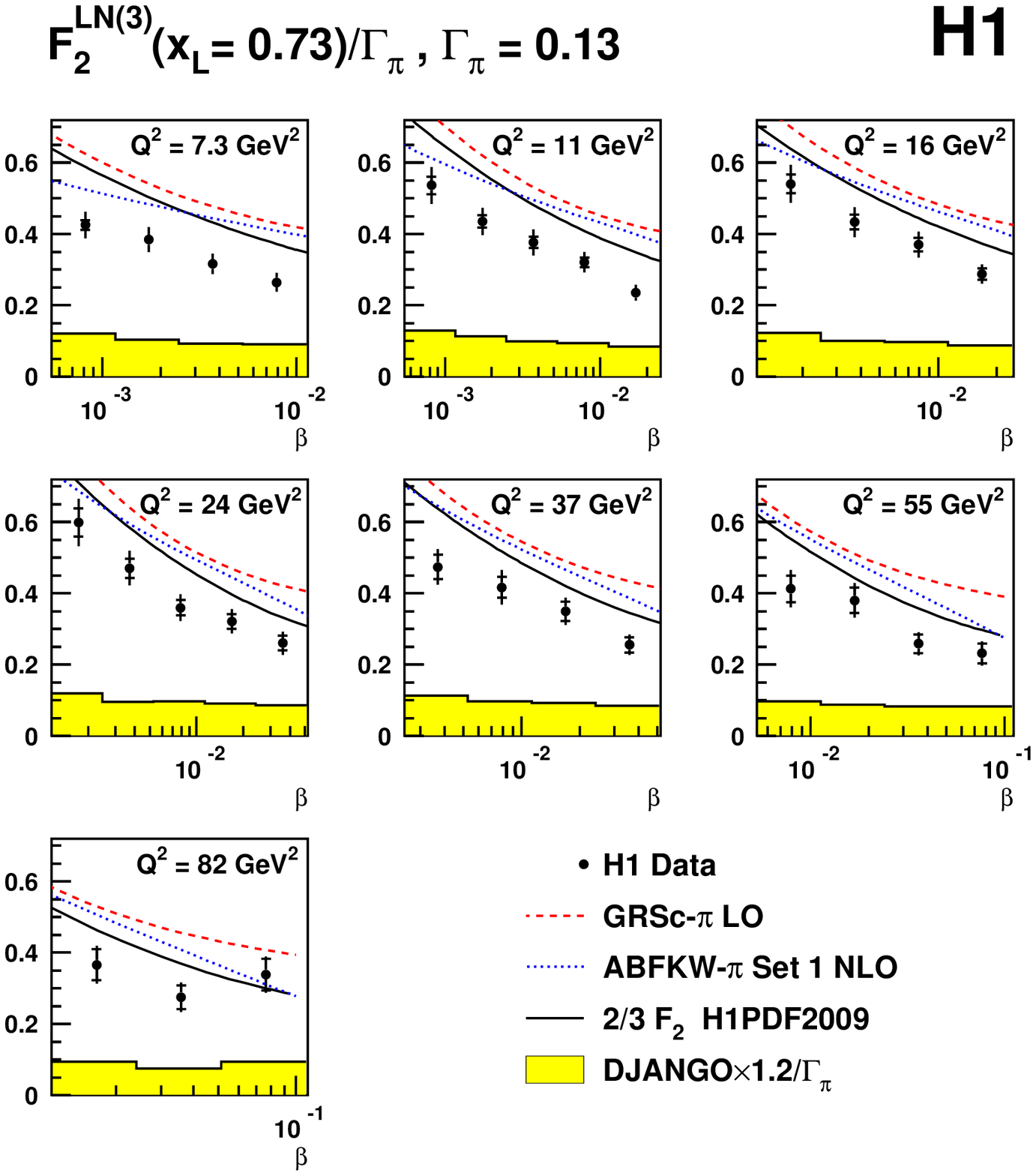,width=170mm}   
\caption{The semi-inclusive structure function $F_2^{LN(3)}$, for
neutrons with $p_T < 0.2~\GeV$, 
divided by the pion flux $\Gamma_\pi$ integrated over
$t$ at the central value $x_L=0.73$, shown
as a function of $\beta$ in bins of $Q^2$.
The pion flux is defined in Eq.~\ref{holtmanflux}. 
The data are compared to two different parameterisations of the
pion structure function $F_2^{\pi}$~\cite{Gluck:1999xe,Aurenche:1989sx}
 and to the H1PDF2009 parameterisation of
the proton structure function \cite{Aaron:2009kv}, which has been scaled by 2/3.
The contribution of neutrons from fragmentation, as predicted by DJANGO 
and scaled by a factor 1.2, 
as described in Sect.~\ref{sec:mc}, is indicated.
}
\label{FNC-F2PI}
\end{figure}

Figure~\ref{FNC-F2PIQ2} shows $F_2^{LN(3)}/\Gamma_{\pi}$ as a
function of $Q^2$ in bins of $\beta$ while
Fig.~\ref{FNC-F2PI} shows $F_2^{LN(3)}/\Gamma_{\pi}$ as a
function of $\beta$ in bins of $Q^2$.
In Fig.~\ref{FNC-F2PI} the contribution of neutrons from fragmentation, 
as predicted by DJANGO and scaled by a weighting factor 1.2, 
as described in Sect.~\ref{sec:mc}, is indicated.
The data are compared to predictions of the parameterisations of the
pion structure function GRSc-$\pi$~\cite{Gluck:1999xe} and 
ABFKW-$\pi$~\cite{Aurenche:1989sx}. 
The measurements are also compared to the H1PDF2009 parameterisation of the
proton structure function $F_2(Q^2,x)$ \cite{Aaron:2009kv} which 
is scaled by a factor of 2/3
in order to naively account for the different number of valence quarks
in the pion and proton, respectively.
The values of $F_2(Q^2,x)$ are calculated at Bjorken-$x$ 
equal to $\beta$.
The $Q^2$ distribution exhibits a rise with increasing $Q^2$ 
(i.e. scaling violation) for all $\beta$ values in the measured range,
which is similar in size and shape to that seen in the parameterisations of the
inclusive structure functions of both the pion and proton 
(Fig.~\ref{FNC-F2PIQ2}).
The $\beta$ distributions show a steep rise
with decreasing  $\beta$ for all $Q^2$ values
(Fig.~\ref{FNC-F2PI}).
This behaviour is in reasonable agreement 
with the pion and proton structure function parameterisations.
In absolute values the parameterisations lie above the measurements.
Other parameterisations of the
pion structure function \cite{Owens:1984zj,Sutton:1991ay} 
were ruled out  by previous H1 and
ZEUS measurements \cite{Adloff:1998yg,Chekanov:2002pf}
as they show a much flatter behaviour as a function of $\beta$.

The comparison of the measured 
$F_2^{LN(3)}/\Gamma_\pi$ and the parameterisations of the
pion structure function is affected by
uncertainties on the pion flux normalisation, as explained above. 
It may also depend on absorptive corrections
\cite{Nikolaev:1997cn,D'Alesio:1998bf,Kaidalov:2006cw,Kopeliovich:2008da}, 
which are not taken into account in this analysis.
Neutron absorption may occur through rescattering which transforms the
neutron into a charged baryon or shifts the neutron to lower 
energy or higher $p_T$.

The results presented here are consistent with the previous
measurement by the H1 Collaboration~\cite{Adloff:1998yg}.
A similar analysis has been published by the ZEUS Collaboration \cite{Chekanov:2002pf}. 
There is good agreement between the two cross section measurements. 
For the extraction of the pion structure function, the flux factor normalisations 
used in the ZEUS analysis are different 
from the one used here.  Within the uncertainties of the normalisation the H1 and ZEUS
results agree.


\section{Summary}

\noindent
The cross section for leading neutron production in deep-inelastic
positron-proton scattering $d\sigma/dx_L$ and the semi-inclusive structure function 
$F_2^{LN(3)}(Q^2,x,x_L)$ are 
measured in the kinematic region
$6< Q^2 < 100~\GeV^2$,
$1.5\cdot 10^{-4} < x < 3\cdot 10^{-2}$, $0.32 < x_L < 0.95$
and $p_T < 0.2~\GeV$.
The present measurements have experimental uncertainties of
$10$ to $15\%$.

The measurements are well described by a
Monte Carlo simulation including neutron production in fragmentation and 
neutrons produced from $\pi^+$ exchange, as predicted by the DJANGO and RAPGAP programs
respectively. 
At large $x_L\gsim 0.7$ the $\pi^+$-exchange process dominates.

Within the measured kinematic range, the semi-inclusive structure 
function $F_2^{LN(3)}$ and the inclusive
structure function $F_2$  have similar $(Q^2,x)$ behaviour,
which is consistent with the hypothesis of limiting fragmentation.
The dependence of $F_2^{LN(3)}$ on the variable $\beta$ is similar
for all $x_L$ bins, in accordance with the expectation from
proton vertex factorisation.
The scaling violations observed in $F_2^{LN(3)}$ are similar in 
size and shape to those seen in the parameterisations of the
inclusive structure functions of the pion and the proton.
The data are used to estimate the structure function of the pion,
up to uncertainties on the background contribution and the 
overall normalisation,
in the framework of the one pion exchange model for the neutron kinematic range $0.68<x_L<0.77$ and $p_T < 0.2~\GeV$.

\section*{Acknowledgements}

We are grateful to the HERA machine group whose outstanding
efforts have made this experiment possible.
We thank the engineers and technicians for their work in constructing and
maintaining the H1 detector, our funding agencies for
financial support, the DESY technical staff for continual assistance
and the DESY directorate for support and for the
hospitality which they extend to the non-DESY members of the 
collaboration.


\begin{thebibliography}{99}

\bibitem{Adloff:1998yg}
C.~Adloff {\em et~al.} [H1 Collaboration],
\newblock Eur. Phys. J. C {\bf 6} (1999) 587 [hep-ex/9811013].

\bibitem{Chekanov:2002pf}
S.~Chekanov {\em et~al.} [ZEUS Collaboration],
\newblock Nucl. Phys. B {\bf 637} (2002) 3 [hep-ex/0205076].

\bibitem{Breitweg:2000nk}
J.~Breitweg {\em et~al.} [ZEUS Collaboration],
\newblock Nucl. Phys. B {\bf 596} (2001) 3 [hep-ex/0010019].

\bibitem{Chekanov:2004wn}
S.~Chekanov {\em et~al.} [ZEUS Collaboration],
\newblock Phys. Lett. B {\bf 610} (2005) 199 [hep-ex/0404002].

\bibitem{Aktas:2004gi}
A.~Aktas {\em et~al.} [H1 Collaboration],
\newblock Eur. Phys. J. C {\bf 41} (2005) 273 [hep-ex/0501074].

\bibitem{Chekanov:2004dk}
S.~Chekanov {\em et~al.} [ZEUS Collaboration],
\newblock Phys. Lett. B {\bf 590} (2004) 143 \mbox{[hep-ex/0401017]}.

\bibitem{Chekanov:2007tv}
S.~Chekanov {\em et~al.} [ZEUS Collaboration],
\newblock Nucl. Phys. B {\bf 776} (2007) 1 [hep-ex/0702028].

\bibitem{Sullivan:1971kd}
J.D.~Sullivan,
\newblock Phys. Rev. D {\bf 5} (1972) 1732.

\bibitem{Bishari:1972tx}
M.~Bishari,
\newblock Phys. Lett. B {\bf 38} (1972) 510.

\bibitem{Holtmann:1994rs}
H.~Holtmann {\em et~al.},
\newblock Phys. Lett. B {\bf 338} (1994) 363.

\bibitem{Kopeliovich:1996iw}
B.~Kopeliovich, B.~Povh and I.~Potashnikova,
\newblock Z. Phys. C {\bf 73} (1996) 125 [hep-ph/9601291].

\bibitem{Przybycien:1996zb}
M.~Przybycie\'n, A.~Szczurek and G.~Ingelman,
Z. Phys. C {\bf 74} (1997) 509 \\
\mbox{[hep-ph/9606294]}.

\bibitem{Szczurek:1997cw}
A.~Szczurek, N.N.~Nikolaev and J.~Speth,
Phys. Lett. B {\bf 428} 383 (1998) 383 \\
\mbox{[hep-ph/9712261]}.

\bibitem{Khoze:2006hw}
V.A.~Khoze, A.D.~Martin and M.G.~Ryskin,
Eur. Phys. J. C {\bf 48} (2006) 797 \\
\mbox{[hep-ph/0606213]}.

\bibitem{Anassontzis:1987hk}
E.~Anassontzis {\em et~al.},
\newblock Phys. Rev. D {\bf 38} (1988) 1377; \newline
J.S.~Conway {\em et~al.},
\newblock Phys. Rev. D {\bf 39} (1989) 92; \newline
J.~Badier {\em et~al.} [NA3 Collaboration],
\newblock Z. Phys. C {\bf 18} (1983) 281; \newline
B.~Betev {\em et~al.} [NA10 Collaboration],
\newblock Z. Phys. C {\bf 28} (1985) 9; \newline
C.De~Marzo {\em et~al.} [NA24 Collaboration],
\newblock Phys. Rev. D {\bf 36} (1987) 8; \newline
M.~Bonesini {\em et~al.} [WA70 Collaboration],
\newblock Z. Phys. C {\bf 37} (1988) 535.

\bibitem{Benecke:1969sh}
J.~Benecke {\em et~al.},
\newblock Phys. Rev. {\bf 188} (1969) 2159.

\bibitem{Chou:1994dh}
T.T.~Chou and C.-N.~Yang,
\newblock Phys. Rev. D {\bf 50} (1994) 590.

\bibitem{Adloff:1997sc}
C.~Adloff {\em et~al.} [H1 Collaboration],
\newblock Z. Phys. C {\bf 76} 613 (1997) 613 [hep-ex/9708016].

\bibitem{Abt:1996hi}
I.~Abt {\em et~al.} [H1 Collaboration],
\newblock Nucl. Instrum. Meth. A {\bf 386} (1997) 310.

\bibitem{Abt:1996xv}
I.~Abt {\em et~al.} [H1 Collaboration],
\newblock Nucl. Instrum. Meth. A {\bf 386} (1997) 348.

\bibitem{Appuhn:1996na}
R.D.~Appuhn {\em et~al.} [H1 SPACAL Group],
\newblock Nucl. Instrum. Meth. A {\bf 386} (1997) 397.

\bibitem{Pitzl:2000wz}
D.~Pitzl {\em et~al.},
\newblock Nucl. Instrum. Meth. A {\bf 454} (2000) 334 [hep-ex/0002044].

\bibitem{Andrieu:1993kh}
B.~Andrieu {\em et~al.} [H1 Calorimeter Group],
\newblock Nucl. Instrum. Meth. A {\bf 336} (1993) 460.

\bibitem{Kleinwort:2006zz}
C.~Kleinwort, 
``H1 Alignment Experience'', 
in Proceedings of the First LHC Detector Alignment Workshop,
eds. S.~Blusk {\it et~al.}, CERN-2007-004, p.41.

\bibitem{Andrieu:1993tz}
B.~Andrieu {\em et~al.} [H1 Calorimeter Group],
\newblock Nucl. Instrum. Meth. A {\bf 336} (1993) 499;
\newline
%
B.~Andrieu {\em et~al.} [H1 Calorimeter Group],
\newblock Nucl. Instrum. Meth. A {\bf 350} (1994) 57.

\bibitem{Aaron:2009kv}
F.D.~Aaron {\em et~al.} [H1 Collaboration], 
DESY-09-005, accepted for publication by EPJC,
arXiv:0904.3513.




%
%
%

\bibitem{Brun:1978fy}
R.~Brun {\em et~al.}, GEANT3,
\newblock CERN-DD/EE/84-1.

\bibitem{Charchula:1994kf}
K.~Charchula, G.A.~Schuler and H.~Spiesberger, DJANGOH~1.4,
\newblock Comput. Phys. Commun. {\bf 81} (1994) 381.

\bibitem{Lonnblad:1992tz}
L.~L\"onnblad, ARIADNE~4.10,
\newblock Comput. Phys. Commun. {\bf 71} (1992) 15.

\bibitem{JETSET}
B.~Andersson {\em et~al.}, JETSET~7.41, Phys. Rept. {\bf 97} (1983) 31.

\bibitem{Jung:1993gf}
H.~Jung, RAPGAP~3.1,
\newblock Comp. Phys. Commun. {\bf 86} (1995) 147.

\bibitem{Kwiatkowski:1990es}
A.~Kwiatkowski, H.~Spiesberger and H.J.~M\"ohring,
\newblock Comp. Phys. Commun. {\bf 69} (1992) 155.

\bibitem{Timmermans:1990tz}
R.G.E.~Timmermans, T.A.~Rijken and J.J.~de~Swart,
\newblock Phys. Rev. Lett. {\bf 67} (1991) 1074.

\bibitem{Gluck:1991ng}
M.~Gl\"uck, E.~Reya and A.~Vogt,
\newblock Z. Phys. C {\bf 53} (1992) 127.

\bibitem{Gluck:1991ey}
M.~Gl\"uck, E.~Reya and A.~Vogt,
\newblock Z. Phys. C {\bf 53} (1992) 651.

\bibitem{PHOJET}
R.~Engel and J.~Ranft, PHOJET~1.0,
\newblock Phys.Rev. D {\bf 54} (1996) 4244 [hep-ph/9509373].

\bibitem{List:1998jz}
B.~List and A.~Mastroberardino, 
``DIFFVM:  A Monte Carlo generator for diffractive processes in  ep scattering'',
in Proceedings of the workshop ``Monte Carlo generators for HERA  physics'',
eds.  A.T.~Doyle {\it et~al.}, DESY-PROC-1999-02, p. 396.

\bibitem{Edin:1995gi}
A.~Edin, G.~Ingelman and J.~Rathsman,
\newblock Phys. Lett. B {\bf 366} (1996) 371 [hep-ph/9508386].

\bibitem{Ingelman:1996mq}
G.~Ingelman, A.~Edin and J.~Rathsman,
\newblock Comput. Phys. Commun. {\bf 101} (1997) 108 [hep-ph/9605286].

\bibitem{Rathsman:1998tp}
J.~Rathsman,
\newblock Phys. Lett. B {\bf 452} (1999) 364 [hep-ph/9812423].

\bibitem{Adloff:2001rw}
C.~Adloff {\em et~al.} [H1 Collaboration],
\newblock Phys. Lett. B {\bf 520} (2001) 183 [hep-ex/0108035].

\bibitem{Gluck:1999xe}
M.~Gl\"uck, E.~Reya and I.~Schienbein,
\newblock Eur. Phys. J. C {\bf 10} (1999) 313 [hep-ph/9903288].

\bibitem{Aurenche:1989sx}
P.~Aurenche {\em et~al.},
\newblock Phys. Lett. B {\bf 233} (1989) 517.

\bibitem{Owens:1984zj}
J.F.~Owens,
\newblock Phys. Rev. D {\bf 30} (1984) 943.

\bibitem{Sutton:1991ay}
P.J.~Sutton {\em et~al.},
\newblock Phys. Rev. D {\bf 45} (1992) 2349.

\bibitem{Nikolaev:1997cn}
N.N.~Nikolaev, J.~Speth and B.G.~Zakharov, 
\mbox{hep-ph/9708290}.

\bibitem{D'Alesio:1998bf}
U.~D'Alesio and H.J.~Pirner,
\newblock Eur. Phys. J. A {\bf 7} (2000) 109 [hep-ph/9806321].

\bibitem{Kaidalov:2006cw}
A.B.~Kaidalov {\em et~al.},
\newblock Eur. Phys. J. C {\bf 47} (2006) 385 [hep-ph/0602215].


\bibitem{Kopeliovich:2008da}
B.~Kopeliovich {\em et~al.},
\newblock Phys. Rev. D {\bf 78} (2008) 014031 [arXiv:0805.4534].

\end{thebibliography}

\begin{table}[p]
\centering
{
\begin{tabular}{|c|c|c|c|c|c|}
\hline
 & & & & & \\[-3mm]
$Q^2$-bin & $Q^2$ range $[\GeV^2]$         & $x$-bin & $x$ range                           & $x_L$-bin & $x_L$ range \\ 
 & & & & & \\[-3mm]
\hline
 & & & & & \\[-3mm]
   $1$    & $6.00\div9.00$ &   $1$   & $1.50\cdot10^{-4}\div3.20\cdot10^{-4}$ & $1$       & $0.32\div0.41$ \\
   $2$    & $9.00\div13.5$ &   $2$   & $3.20\cdot10^{-4}\div6.82\cdot10^{-4}$ & $2$       & $0.41\div0.50$ \\
   $3$    & $13.5\div20.0$ &   $3$   & $6.82\cdot10^{-4}\div1.45\cdot10^{-3}$ & $3$       & $0.50\div0.59$ \\
   $4$    & $20.0\div30.0$ &   $4$   & $1.45\cdot10^{-3}\div3.10\cdot10^{-3}$ & $4$       & $0.59\div0.68$ \\
   $5$    & $30.0\div45.0$ &   $5$   & $3.10\cdot10^{-3}\div6.60\cdot10^{-3}$ & $5$       & $0.68\div0.77$ \\
   $6$    & $45.0\div67.0$ &   $6$   & $6.60\cdot10^{-3}\div1.41\cdot10^{-2}$ & $6$       & $0.77\div0.86$ \\
   $7$    & $67.0\div100 $ &   $7$   & $1.41\cdot10^{-2}\div3.00\cdot10^{-2}$ & $7$       & $0.86\div0.95$ \\
\hline
\end{tabular}
}
\caption{Bins in $Q^2$, $x$ and $x_L$ as used for the measurement of 
the semi-inclusive structure function $F_2^{LN(3)}$.}
\label{tab:binlim}
\end{table}

\def\tabcaptxtadd{The first uncertainty is statistical and the second  
systematic. Normalisation uncertainties of $5\%$ are not included.}

\begin{table}[p]
\centering
{
\begin{tabular}{|c|r|r|}
\hline
 & & \\[-3mm]
$x_L$ range & ${d\sigma}/{d x_L} ~[\mathrm{nb}]\hspace*{3.5ex}$ & ${d\sigma}/{d x_L} 
~[\mathrm{nb}]\hspace*{4ex}$  \\[3pt]
 & $p_T< x_L\cdot0.69~\GeV$  & $p_T<0.2~\GeV~~~~~~$  \\ 
 & & \\[-3mm]
\hline
 & & \\[-4mm]
$0.30\div0.37$ & $10.6 ~\pm~ 0.11 ~\pm~ 1.3$ & $8.21 ~\pm~ 0.09 ~\pm~ 0.99$ \\
$0.37\div0.44$ & $12.0 ~\pm~ 0.11 ~\pm~ 1.1$ & $6.75 ~\pm~ 0.07 ~\pm~ 0.53$ \\
$0.44\div0.51$ & $14.0 ~\pm~ 0.11 ~\pm~ 1.3$ & $6.26 ~\pm~ 0.06 ~\pm~ 0.45$ \\
$0.51\div0.58$ & $15.6 ~\pm~ 0.11 ~\pm~ 1.4$ & $5.92 ~\pm~ 0.05 ~\pm~ 0.46$ \\
$0.58\div0.65$ & $16.5 ~\pm~ 0.11 ~\pm~ 1.4$ & $5.81 ~\pm~ 0.05 ~\pm~ 0.42$ \\
$0.65\div0.72$ & $16.7 ~\pm~ 0.10 ~\pm~ 1.4$ & $5.71 ~\pm~ 0.05 ~\pm~ 0.40$ \\
$0.72\div0.79$ & $16.0 ~\pm~ 0.10 ~\pm~ 1.3$ & $5.57 ~\pm~ 0.05 ~\pm~ 0.31$ \\
$0.79\div0.86$ & $12.4 ~\pm~ 0.08 ~\pm~ 1.2$ & $4.54 ~\pm~ 0.04 ~\pm~ 0.25$ \\
$0.86\div0.93$ & $7.3 ~\pm~ 0.06 ~\pm~ 1.2$ & $2.57 ~\pm~ 0.03 ~\pm~ 0.38$ \\
$0.93\div1.00$ & $2.3 ~\pm~ 0.03 ~\pm~ 0.9$ & $0.63 ~\pm~ 0.01 ~\pm~ 0.26$ \\
\hline
\end{tabular}
}
\caption{Differential cross section $\der\sigma/{\der x_L}$ of leading neutron production
 in deep-inelastic scattering in the kinematic range
 $6 < Q^2 < 100~\GeV^2$, $1.5\cdot 10^{-4}
 < x < 3\cdot 10^{-2}$ and $0.32 < x_L < 0.95$.
 \tabcaptxtadd
}
\label{tab:dsdxl1}  
\end{table}

\begin{table}[p]
\centering
\scalebox{1.25}{
{\tiny
\begin{tabular}{ccccc}
\hline
 & & & & \\[-1mm]
$Q^2 ~[\mathrm{GeV}^2]$  &  $x$  &  $x_L$  &  $\beta$  &  $F_2^{LN(3)}$  \\ 
 & & & & \\[-1mm]
\hline
 & & & & \\[-1mm] 
$7.3$ & $2.24\cdot10^{-4}$ & $0.365$ & $3.53\cdot10^{-4}$ & $0.0724 ~\pm~ 0.0035 ~\pm~ 0.0058$ \\
$7.3$ & $2.24\cdot10^{-4}$ & $0.455$ & $4.12\cdot10^{-4}$ & $0.0573 ~\pm~ 0.0024 ~\pm~ 0.0044$ \\
$7.3$ & $2.24\cdot10^{-4}$ & $0.545$ & $4.93\cdot10^{-4}$ & $0.0582 ~\pm~ 0.0022 ~\pm~ 0.0049$ \\
$7.3$ & $2.24\cdot10^{-4}$ & $0.635$ & $6.14\cdot10^{-4}$ & $0.0589 ~\pm~ 0.0020 ~\pm~ 0.0042$ \\
$7.3$ & $2.24\cdot10^{-4}$ & $0.725$ & $8.16\cdot10^{-4}$ & $0.0557 ~\pm~ 0.0018 ~\pm~ 0.0036$ \\
$7.3$ & $2.24\cdot10^{-4}$ & $0.815$ & $1.21\cdot10^{-3}$ & $0.0489 ~\pm~ 0.0017 ~\pm~ 0.0023$ \\
$7.3$ & $2.24\cdot10^{-4}$ & $0.905$ & $2.36\cdot10^{-3}$ & $0.0236 ~\pm~ 0.0012 ~\pm~ 0.0043$ \\
 & & & & \\[-1mm] 
$7.3$ & $4.78\cdot10^{-4}$ & $0.365$ & $7.53\cdot10^{-4}$ & $0.0627 ~\pm~ 0.0030 ~\pm~ 0.0052$ \\
$7.3$ & $4.78\cdot10^{-4}$ & $0.455$ & $8.77\cdot10^{-4}$ & $0.0547 ~\pm~ 0.0022 ~\pm~ 0.0037$ \\
$7.3$ & $4.78\cdot10^{-4}$ & $0.545$ & $1.05\cdot10^{-3}$ & $0.0490 ~\pm~ 0.0018 ~\pm~ 0.0046$ \\
$7.3$ & $4.78\cdot10^{-4}$ & $0.635$ & $1.31\cdot10^{-3}$ & $0.0500 ~\pm~ 0.0017 ~\pm~ 0.0034$ \\
$7.3$ & $4.78\cdot10^{-4}$ & $0.725$ & $1.74\cdot10^{-3}$ & $0.0503 ~\pm~ 0.0016 ~\pm~ 0.0034$ \\
$7.3$ & $4.78\cdot10^{-4}$ & $0.815$ & $2.58\cdot10^{-3}$ & $0.0425 ~\pm~ 0.0015 ~\pm~ 0.0021$ \\
$7.3$ & $4.78\cdot10^{-4}$ & $0.905$ & $5.03\cdot10^{-3}$ & $0.0205 ~\pm~ 0.0010 ~\pm~ 0.0030$ \\
 & & & & \\[-1mm] 
$7.3$ & $1.02\cdot10^{-3}$ & $0.365$ & $1.60\cdot10^{-3}$ & $0.0531 ~\pm~ 0.0026 ~\pm~ 0.0045$ \\
$7.3$ & $1.02\cdot10^{-3}$ & $0.455$ & $1.87\cdot10^{-3}$ & $0.0471 ~\pm~ 0.0020 ~\pm~ 0.0036$ \\
$7.3$ & $1.02\cdot10^{-3}$ & $0.545$ & $2.24\cdot10^{-3}$ & $0.0421 ~\pm~ 0.0016 ~\pm~ 0.0031$ \\
$7.3$ & $1.02\cdot10^{-3}$ & $0.635$ & $2.79\cdot10^{-3}$ & $0.0445 ~\pm~ 0.0015 ~\pm~ 0.0030$ \\
$7.3$ & $1.02\cdot10^{-3}$ & $0.725$ & $3.71\cdot10^{-3}$ & $0.0413 ~\pm~ 0.0014 ~\pm~ 0.0028$ \\
$7.3$ & $1.02\cdot10^{-3}$ & $0.815$ & $5.51\cdot10^{-3}$ & $0.0352 ~\pm~ 0.0013 ~\pm~ 0.0019$ \\
$7.3$ & $1.02\cdot10^{-3}$ & $0.905$ & $1.07\cdot10^{-2}$ & $0.0156 ~\pm~ 0.0008 ~\pm~ 0.0023$ \\
 & & & & \\[-1mm] 
$7.3$ & $2.17\cdot10^{-3}$ & $0.365$ & $3.42\cdot10^{-3}$ & $0.0461 ~\pm~ 0.0025 ~\pm~ 0.0037$ \\
$7.3$ & $2.17\cdot10^{-3}$ & $0.455$ & $3.99\cdot10^{-3}$ & $0.0385 ~\pm~ 0.0018 ~\pm~ 0.0028$ \\
$7.3$ & $2.17\cdot10^{-3}$ & $0.545$ & $4.77\cdot10^{-3}$ & $0.0383 ~\pm~ 0.0016 ~\pm~ 0.0030$ \\
$7.3$ & $2.17\cdot10^{-3}$ & $0.635$ & $5.95\cdot10^{-3}$ & $0.0338 ~\pm~ 0.0013 ~\pm~ 0.0021$ \\
$7.3$ & $2.17\cdot10^{-3}$ & $0.725$ & $7.90\cdot10^{-3}$ & $0.0346 ~\pm~ 0.0013 ~\pm~ 0.0026$ \\
$7.3$ & $2.17\cdot10^{-3}$ & $0.815$ & $1.17\cdot10^{-2}$ & $0.0293 ~\pm~ 0.0011 ~\pm~ 0.0015$ \\
$7.3$ & $2.17\cdot10^{-3}$ & $0.905$ & $2.29\cdot10^{-2}$ & $0.0147 ~\pm~ 0.0008 ~\pm~ 0.0024$ \\
 & & & & \\[-1mm] 
$ 11$ & $2.24\cdot10^{-4}$ & $0.365$ & $3.53\cdot10^{-4}$ & $0.0866 ~\pm~ 0.0058 ~\pm~ 0.0069$ \\
$ 11$ & $2.24\cdot10^{-4}$ & $0.455$ & $4.12\cdot10^{-4}$ & $0.0740 ~\pm~ 0.0042 ~\pm~ 0.0060$ \\
$ 11$ & $2.24\cdot10^{-4}$ & $0.545$ & $4.93\cdot10^{-4}$ & $0.0692 ~\pm~ 0.0036 ~\pm~ 0.0049$ \\
$ 11$ & $2.24\cdot10^{-4}$ & $0.635$ & $6.14\cdot10^{-4}$ & $0.0701 ~\pm~ 0.0033 ~\pm~ 0.0052$ \\
$ 11$ & $2.24\cdot10^{-4}$ & $0.725$ & $8.16\cdot10^{-4}$ & $0.0702 ~\pm~ 0.0031 ~\pm~ 0.0049$ \\
$ 11$ & $2.24\cdot10^{-4}$ & $0.815$ & $1.21\cdot10^{-3}$ & $0.0517 ~\pm~ 0.0025 ~\pm~ 0.0028$ \\
$ 11$ & $2.24\cdot10^{-4}$ & $0.905$ & $2.36\cdot10^{-3}$ & $0.0273 ~\pm~ 0.0019 ~\pm~ 0.0046$ \\
 & & & & \\[-1mm] 
$ 11$ & $4.78\cdot10^{-4}$ & $0.365$ & $7.53\cdot10^{-4}$ & $0.0771 ~\pm~ 0.0044 ~\pm~ 0.0061$ \\
$ 11$ & $4.78\cdot10^{-4}$ & $0.455$ & $8.77\cdot10^{-4}$ & $0.0652 ~\pm~ 0.0032 ~\pm~ 0.0047$ \\
$ 11$ & $4.78\cdot10^{-4}$ & $0.545$ & $1.05\cdot10^{-3}$ & $0.0580 ~\pm~ 0.0026 ~\pm~ 0.0051$ \\
$ 11$ & $4.78\cdot10^{-4}$ & $0.635$ & $1.31\cdot10^{-3}$ & $0.0635 ~\pm~ 0.0025 ~\pm~ 0.0046$ \\
$ 11$ & $4.78\cdot10^{-4}$ & $0.725$ & $1.74\cdot10^{-3}$ & $0.0570 ~\pm~ 0.0022 ~\pm~ 0.0034$ \\
$ 11$ & $4.78\cdot10^{-4}$ & $0.815$ & $2.58\cdot10^{-3}$ & $0.0483 ~\pm~ 0.0020 ~\pm~ 0.0023$ \\
$ 11$ & $4.78\cdot10^{-4}$ & $0.905$ & $5.03\cdot10^{-3}$ & $0.0226 ~\pm~ 0.0014 ~\pm~ 0.0039$ \\
 & & & & \\[-1mm] 
$ 11$ & $1.02\cdot10^{-3}$ & $0.365$ & $1.60\cdot10^{-3}$ & $0.0623 ~\pm~ 0.0037 ~\pm~ 0.0052$ \\
$ 11$ & $1.02\cdot10^{-3}$ & $0.455$ & $1.87\cdot10^{-3}$ & $0.0554 ~\pm~ 0.0028 ~\pm~ 0.0041$ \\
$ 11$ & $1.02\cdot10^{-3}$ & $0.545$ & $2.24\cdot10^{-3}$ & $0.0513 ~\pm~ 0.0023 ~\pm~ 0.0042$ \\
$ 11$ & $1.02\cdot10^{-3}$ & $0.635$ & $2.79\cdot10^{-3}$ & $0.0483 ~\pm~ 0.0020 ~\pm~ 0.0030$ \\
$ 11$ & $1.02\cdot10^{-3}$ & $0.725$ & $3.71\cdot10^{-3}$ & $0.0493 ~\pm~ 0.0020 ~\pm~ 0.0035$ \\
$ 11$ & $1.02\cdot10^{-3}$ & $0.815$ & $5.51\cdot10^{-3}$ & $0.0419 ~\pm~ 0.0018 ~\pm~ 0.0021$ \\
$ 11$ & $1.02\cdot10^{-3}$ & $0.905$ & $1.07\cdot10^{-2}$ & $0.0182 ~\pm~ 0.0011 ~\pm~ 0.0033$ \\
 & & & & \\[-1mm] 
\hline
\end{tabular}
}
}
\caption{The semi-inclusive structure function $F_2^{LN(3)}(Q^2,x,x_L)$, for
 neutrons with \mbox{$p_T < 0.2~\GeV$}.
  The bin centre values of $Q^2$, $x$, $x_L$ and $\beta$
  in the corresponding bins are also given.
 \tabcaptxtadd}
\label{tab:f2ln3p1}
\end{table}

\begin{table}[p]
\centering
\scalebox{1.25}{
{\tiny
\begin{tabular}{ccccc}
\hline
 & & & & \\[-1mm]
$Q^2 ~[\mathrm{GeV}^2]$  &  $x$  &  $x_L$  &  $\beta$  &  $F_2^{LN(3)}$  \\ 
 & & & & \\[-1mm]
\hline
 & & & & \\[-1mm] 
$ 11$ & $2.17\cdot10^{-3}$ & $0.365$ & $3.42\cdot10^{-3}$ & $0.0505 ~\pm~ 0.0032 ~\pm~ 0.0041$ \\
$ 11$ & $2.17\cdot10^{-3}$ & $0.455$ & $3.99\cdot10^{-3}$ & $0.0447 ~\pm~ 0.0024 ~\pm~ 0.0035$ \\
$ 11$ & $2.17\cdot10^{-3}$ & $0.545$ & $4.77\cdot10^{-3}$ & $0.0428 ~\pm~ 0.0021 ~\pm~ 0.0032$ \\
$ 11$ & $2.17\cdot10^{-3}$ & $0.635$ & $5.95\cdot10^{-3}$ & $0.0417 ~\pm~ 0.0018 ~\pm~ 0.0034$ \\
$ 11$ & $2.17\cdot10^{-3}$ & $0.725$ & $7.90\cdot10^{-3}$ & $0.0420 ~\pm~ 0.0018 ~\pm~ 0.0026$ \\
$ 11$ & $2.17\cdot10^{-3}$ & $0.815$ & $1.17\cdot10^{-2}$ & $0.0339 ~\pm~ 0.0016 ~\pm~ 0.0016$ \\
$ 11$ & $2.17\cdot10^{-3}$ & $0.905$ & $2.29\cdot10^{-2}$ & $0.0167 ~\pm~ 0.0011 ~\pm~ 0.0031$ \\
 & & & & \\[-1mm] 
$ 11$ & $4.63\cdot10^{-3}$ & $0.365$ & $7.29\cdot10^{-3}$ & $0.0460 ~\pm~ 0.0035 ~\pm~ 0.0041$ \\
$ 11$ & $4.63\cdot10^{-3}$ & $0.455$ & $8.50\cdot10^{-3}$ & $0.0412 ~\pm~ 0.0027 ~\pm~ 0.0032$ \\
$ 11$ & $4.63\cdot10^{-3}$ & $0.545$ & $1.02\cdot10^{-2}$ & $0.0348 ~\pm~ 0.0021 ~\pm~ 0.0027$ \\
$ 11$ & $4.63\cdot10^{-3}$ & $0.635$ & $1.27\cdot10^{-2}$ & $0.0344 ~\pm~ 0.0019 ~\pm~ 0.0025$ \\
$ 11$ & $4.63\cdot10^{-3}$ & $0.725$ & $1.68\cdot10^{-2}$ & $0.0307 ~\pm~ 0.0016 ~\pm~ 0.0018$ \\
$ 11$ & $4.63\cdot10^{-3}$ & $0.815$ & $2.50\cdot10^{-2}$ & $0.0269 ~\pm~ 0.0015 ~\pm~ 0.0012$ \\
$ 11$ & $4.63\cdot10^{-3}$ & $0.905$ & $4.87\cdot10^{-2}$ & $0.0157 ~\pm~ 0.0012 ~\pm~ 0.0023$ \\
 & & & & \\[-1mm] 
$ 16$ & $4.78\cdot10^{-4}$ & $0.365$ & $7.53\cdot10^{-4}$ & $0.0829 ~\pm~ 0.0059 ~\pm~ 0.0069$ \\
$ 16$ & $4.78\cdot10^{-4}$ & $0.455$ & $8.77\cdot10^{-4}$ & $0.0652 ~\pm~ 0.0039 ~\pm~ 0.0049$ \\
$ 16$ & $4.78\cdot10^{-4}$ & $0.545$ & $1.05\cdot10^{-3}$ & $0.0679 ~\pm~ 0.0037 ~\pm~ 0.0049$ \\
$ 16$ & $4.78\cdot10^{-4}$ & $0.635$ & $1.31\cdot10^{-3}$ & $0.0666 ~\pm~ 0.0033 ~\pm~ 0.0052$ \\
$ 16$ & $4.78\cdot10^{-4}$ & $0.725$ & $1.74\cdot10^{-3}$ & $0.0708 ~\pm~ 0.0034 ~\pm~ 0.0049$ \\
$ 16$ & $4.78\cdot10^{-4}$ & $0.815$ & $2.58\cdot10^{-3}$ & $0.0578 ~\pm~ 0.0030 ~\pm~ 0.0030$ \\
$ 16$ & $4.78\cdot10^{-4}$ & $0.905$ & $5.03\cdot10^{-3}$ & $0.0300 ~\pm~ 0.0022 ~\pm~ 0.0051$ \\
 & & & & \\[-1mm] 
$ 16$ & $1.02\cdot10^{-3}$ & $0.365$ & $1.60\cdot10^{-3}$ & $0.0787 ~\pm~ 0.0054 ~\pm~ 0.0068$ \\
$ 16$ & $1.02\cdot10^{-3}$ & $0.455$ & $1.87\cdot10^{-3}$ & $0.0706 ~\pm~ 0.0043 ~\pm~ 0.0050$ \\
$ 16$ & $1.02\cdot10^{-3}$ & $0.545$ & $2.24\cdot10^{-3}$ & $0.0529 ~\pm~ 0.0029 ~\pm~ 0.0040$ \\
$ 16$ & $1.02\cdot10^{-3}$ & $0.635$ & $2.79\cdot10^{-3}$ & $0.0580 ~\pm~ 0.0029 ~\pm~ 0.0043$ \\
$ 16$ & $1.02\cdot10^{-3}$ & $0.725$ & $3.71\cdot10^{-3}$ & $0.0568 ~\pm~ 0.0028 ~\pm~ 0.0038$ \\
$ 16$ & $1.02\cdot10^{-3}$ & $0.815$ & $5.51\cdot10^{-3}$ & $0.0469 ~\pm~ 0.0024 ~\pm~ 0.0023$ \\
$ 16$ & $1.02\cdot10^{-3}$ & $0.905$ & $1.07\cdot10^{-2}$ & $0.0233 ~\pm~ 0.0017 ~\pm~ 0.0041$ \\
 & & & & \\[-1mm] 
$ 16$ & $2.17\cdot10^{-3}$ & $0.365$ & $3.42\cdot10^{-3}$ & $0.0674 ~\pm~ 0.0050 ~\pm~ 0.0055$ \\
$ 16$ & $2.17\cdot10^{-3}$ & $0.455$ & $3.99\cdot10^{-3}$ & $0.0546 ~\pm~ 0.0035 ~\pm~ 0.0036$ \\
$ 16$ & $2.17\cdot10^{-3}$ & $0.545$ & $4.77\cdot10^{-3}$ & $0.0512 ~\pm~ 0.0030 ~\pm~ 0.0039$ \\
$ 16$ & $2.17\cdot10^{-3}$ & $0.635$ & $5.95\cdot10^{-3}$ & $0.0513 ~\pm~ 0.0027 ~\pm~ 0.0042$ \\
$ 16$ & $2.17\cdot10^{-3}$ & $0.725$ & $7.90\cdot10^{-3}$ & $0.0485 ~\pm~ 0.0025 ~\pm~ 0.0032$ \\
$ 16$ & $2.17\cdot10^{-3}$ & $0.815$ & $1.17\cdot10^{-2}$ & $0.0416 ~\pm~ 0.0023 ~\pm~ 0.0021$ \\
$ 16$ & $2.17\cdot10^{-3}$ & $0.905$ & $2.29\cdot10^{-2}$ & $0.0195 ~\pm~ 0.0015 ~\pm~ 0.0036$ \\
 & & & & \\[-1mm] 
$ 16$ & $4.63\cdot10^{-3}$ & $0.365$ & $7.29\cdot10^{-3}$ & $0.0544 ~\pm~ 0.0045 ~\pm~ 0.0039$ \\
$ 16$ & $4.63\cdot10^{-3}$ & $0.455$ & $8.50\cdot10^{-3}$ & $0.0425 ~\pm~ 0.0029 ~\pm~ 0.0030$ \\
$ 16$ & $4.63\cdot10^{-3}$ & $0.545$ & $1.02\cdot10^{-2}$ & $0.0417 ~\pm~ 0.0026 ~\pm~ 0.0032$ \\
$ 16$ & $4.63\cdot10^{-3}$ & $0.635$ & $1.27\cdot10^{-2}$ & $0.0421 ~\pm~ 0.0024 ~\pm~ 0.0037$ \\
$ 16$ & $4.63\cdot10^{-3}$ & $0.725$ & $1.68\cdot10^{-2}$ & $0.0377 ~\pm~ 0.0021 ~\pm~ 0.0021$ \\
$ 16$ & $4.63\cdot10^{-3}$ & $0.815$ & $2.50\cdot10^{-2}$ & $0.0307 ~\pm~ 0.0018 ~\pm~ 0.0017$ \\
$ 16$ & $4.63\cdot10^{-3}$ & $0.905$ & $4.87\cdot10^{-2}$ & $0.0151 ~\pm~ 0.0012 ~\pm~ 0.0023$ \\
 & & & & \\[-1mm] 
$ 24$ & $4.78\cdot10^{-4}$ & $0.365$ & $7.53\cdot10^{-4}$ & $0.0945 ~\pm~ 0.0091 ~\pm~ 0.0076$ \\
$ 24$ & $4.78\cdot10^{-4}$ & $0.455$ & $8.77\cdot10^{-4}$ & $0.0903 ~\pm~ 0.0076 ~\pm~ 0.0075$ \\
$ 24$ & $4.78\cdot10^{-4}$ & $0.545$ & $1.05\cdot10^{-3}$ & $0.0716 ~\pm~ 0.0054 ~\pm~ 0.0047$ \\
$ 24$ & $4.78\cdot10^{-4}$ & $0.635$ & $1.31\cdot10^{-3}$ & $0.0664 ~\pm~ 0.0047 ~\pm~ 0.0049$ \\
$ 24$ & $4.78\cdot10^{-4}$ & $0.725$ & $1.74\cdot10^{-3}$ & $0.0784 ~\pm~ 0.0052 ~\pm~ 0.0059$ \\
$ 24$ & $4.78\cdot10^{-4}$ & $0.815$ & $2.58\cdot10^{-3}$ & $0.0640 ~\pm~ 0.0047 ~\pm~ 0.0032$ \\
$ 24$ & $4.78\cdot10^{-4}$ & $0.905$ & $5.03\cdot10^{-3}$ & $0.0281 ~\pm~ 0.0029 ~\pm~ 0.0062$ \\
 & & & & \\[-1mm] 
\hline
\end{tabular}
}
}
\caption{The semi-inclusive structure function $F_2^{LN(3)}(Q^2,x,x_L)$, for
neutrons with \mbox{$p_T < 0.2~\GeV$},  continued from Table~\ref{tab:f2ln3p1}.}
\label{tab:f2ln3p2}
\end{table}

\begin{table}[p]
\centering
\scalebox{1.25}{
{\tiny
\begin{tabular}{ccccc}
\hline
 & & & & \\[-1mm]
$Q^2 ~[\mathrm{GeV}^2]$  &  $x$  &  $x_L$  &  $\beta$  &  $F_2^{LN(3)}$  \\ 
 & & & & \\[-1mm]
\hline
 & & & & \\[-1mm] 
$ 24$ & $1.02\cdot10^{-3}$ & $0.365$ & $1.60\cdot10^{-3}$ & $0.0871 ~\pm~ 0.0073 ~\pm~ 0.0070$ \\
$ 24$ & $1.02\cdot10^{-3}$ & $0.455$ & $1.87\cdot10^{-3}$ & $0.0782 ~\pm~ 0.0056 ~\pm~ 0.0064$ \\
$ 24$ & $1.02\cdot10^{-3}$ & $0.545$ & $2.24\cdot10^{-3}$ & $0.0679 ~\pm~ 0.0044 ~\pm~ 0.0040$ \\
$ 24$ & $1.02\cdot10^{-3}$ & $0.635$ & $2.79\cdot10^{-3}$ & $0.0678 ~\pm~ 0.0040 ~\pm~ 0.0054$ \\
$ 24$ & $1.02\cdot10^{-3}$ & $0.725$ & $3.71\cdot10^{-3}$ & $0.0617 ~\pm~ 0.0035 ~\pm~ 0.0044$ \\
$ 24$ & $1.02\cdot10^{-3}$ & $0.815$ & $5.51\cdot10^{-3}$ & $0.0494 ~\pm~ 0.0031 ~\pm~ 0.0028$ \\
$ 24$ & $1.02\cdot10^{-3}$ & $0.905$ & $1.07\cdot10^{-2}$ & $0.0231 ~\pm~ 0.0021 ~\pm~ 0.0036$ \\
 & & & & \\[-1mm] 
$ 24$ & $2.17\cdot10^{-3}$ & $0.365$ & $3.42\cdot10^{-3}$ & $0.0684 ~\pm~ 0.0060 ~\pm~ 0.0058$ \\
$ 24$ & $2.17\cdot10^{-3}$ & $0.455$ & $3.99\cdot10^{-3}$ & $0.0602 ~\pm~ 0.0045 ~\pm~ 0.0046$ \\
$ 24$ & $2.17\cdot10^{-3}$ & $0.545$ & $4.77\cdot10^{-3}$ & $0.0601 ~\pm~ 0.0041 ~\pm~ 0.0057$ \\
$ 24$ & $2.17\cdot10^{-3}$ & $0.635$ & $5.95\cdot10^{-3}$ & $0.0545 ~\pm~ 0.0034 ~\pm~ 0.0028$ \\
$ 24$ & $2.17\cdot10^{-3}$ & $0.725$ & $7.90\cdot10^{-3}$ & $0.0471 ~\pm~ 0.0029 ~\pm~ 0.0031$ \\
$ 24$ & $2.17\cdot10^{-3}$ & $0.815$ & $1.17\cdot10^{-2}$ & $0.0429 ~\pm~ 0.0028 ~\pm~ 0.0020$ \\
$ 24$ & $2.17\cdot10^{-3}$ & $0.905$ & $2.29\cdot10^{-2}$ & $0.0206 ~\pm~ 0.0019 ~\pm~ 0.0039$ \\
 & & & & \\[-1mm] 
$ 24$ & $4.63\cdot10^{-3}$ & $0.365$ & $7.29\cdot10^{-3}$ & $0.0531 ~\pm~ 0.0050 ~\pm~ 0.0041$ \\
$ 24$ & $4.63\cdot10^{-3}$ & $0.455$ & $8.50\cdot10^{-3}$ & $0.0503 ~\pm~ 0.0040 ~\pm~ 0.0031$ \\
$ 24$ & $4.63\cdot10^{-3}$ & $0.545$ & $1.02\cdot10^{-2}$ & $0.0467 ~\pm~ 0.0034 ~\pm~ 0.0038$ \\
$ 24$ & $4.63\cdot10^{-3}$ & $0.635$ & $1.27\cdot10^{-2}$ & $0.0426 ~\pm~ 0.0028 ~\pm~ 0.0033$ \\
$ 24$ & $4.63\cdot10^{-3}$ & $0.725$ & $1.68\cdot10^{-2}$ & $0.0421 ~\pm~ 0.0027 ~\pm~ 0.0030$ \\
$ 24$ & $4.63\cdot10^{-3}$ & $0.815$ & $2.50\cdot10^{-2}$ & $0.0328 ~\pm~ 0.0022 ~\pm~ 0.0019$ \\
$ 24$ & $4.63\cdot10^{-3}$ & $0.905$ & $4.87\cdot10^{-2}$ & $0.0174 ~\pm~ 0.0017 ~\pm~ 0.0028$ \\
 & & & & \\[-1mm] 
$ 24$ & $9.87\cdot10^{-3}$ & $0.365$ & $1.55\cdot10^{-2}$ & $0.0452 ~\pm~ 0.0052 ~\pm~ 0.0035$ \\
$ 24$ & $9.87\cdot10^{-3}$ & $0.455$ & $1.81\cdot10^{-2}$ & $0.0421 ~\pm~ 0.0042 ~\pm~ 0.0034$ \\
$ 24$ & $9.87\cdot10^{-3}$ & $0.545$ & $2.17\cdot10^{-2}$ & $0.0398 ~\pm~ 0.0035 ~\pm~ 0.0038$ \\
$ 24$ & $9.87\cdot10^{-3}$ & $0.635$ & $2.70\cdot10^{-2}$ & $0.0345 ~\pm~ 0.0028 ~\pm~ 0.0018$ \\
$ 24$ & $9.87\cdot10^{-3}$ & $0.725$ & $3.59\cdot10^{-2}$ & $0.0341 ~\pm~ 0.0027 ~\pm~ 0.0028$ \\
$ 24$ & $9.87\cdot10^{-3}$ & $0.815$ & $5.34\cdot10^{-2}$ & $0.0266 ~\pm~ 0.0022 ~\pm~ 0.0017$ \\
$ 24$ & $9.87\cdot10^{-3}$ & $0.905$ & $1.04\cdot10^{-1}$ & $0.0143 ~\pm~ 0.0015 ~\pm~ 0.0029$ \\
 & & & & \\[-1mm] 
$ 37$ & $1.02\cdot10^{-3}$ & $0.365$ & $1.60\cdot10^{-3}$ & $0.1051 ~\pm~ 0.0109 ~\pm~ 0.0088$ \\
$ 37$ & $1.02\cdot10^{-3}$ & $0.455$ & $1.87\cdot10^{-3}$ & $0.0708 ~\pm~ 0.0063 ~\pm~ 0.0053$ \\
$ 37$ & $1.02\cdot10^{-3}$ & $0.545$ & $2.24\cdot10^{-3}$ & $0.0726 ~\pm~ 0.0058 ~\pm~ 0.0054$ \\
$ 37$ & $1.02\cdot10^{-3}$ & $0.635$ & $2.79\cdot10^{-3}$ & $0.0720 ~\pm~ 0.0052 ~\pm~ 0.0056$ \\
$ 37$ & $1.02\cdot10^{-3}$ & $0.725$ & $3.71\cdot10^{-3}$ & $0.0621 ~\pm~ 0.0045 ~\pm~ 0.0036$ \\
$ 37$ & $1.02\cdot10^{-3}$ & $0.815$ & $5.51\cdot10^{-3}$ & $0.0497 ~\pm~ 0.0039 ~\pm~ 0.0025$ \\
$ 37$ & $1.02\cdot10^{-3}$ & $0.905$ & $1.07\cdot10^{-2}$ & $0.0236 ~\pm~ 0.0026 ~\pm~ 0.0044$ \\
 & & & & \\[-1mm] 
$ 37$ & $2.17\cdot10^{-3}$ & $0.365$ & $3.42\cdot10^{-3}$ & $0.0757 ~\pm~ 0.0081 ~\pm~ 0.0065$ \\
$ 37$ & $2.17\cdot10^{-3}$ & $0.455$ & $3.99\cdot10^{-3}$ & $0.0667 ~\pm~ 0.0060 ~\pm~ 0.0041$ \\
$ 37$ & $2.17\cdot10^{-3}$ & $0.545$ & $4.77\cdot10^{-3}$ & $0.0571 ~\pm~ 0.0046 ~\pm~ 0.0045$ \\
$ 37$ & $2.17\cdot10^{-3}$ & $0.635$ & $5.95\cdot10^{-3}$ & $0.0593 ~\pm~ 0.0044 ~\pm~ 0.0042$ \\
$ 37$ & $2.17\cdot10^{-3}$ & $0.725$ & $7.90\cdot10^{-3}$ & $0.0546 ~\pm~ 0.0039 ~\pm~ 0.0041$ \\
$ 37$ & $2.17\cdot10^{-3}$ & $0.815$ & $1.17\cdot10^{-2}$ & $0.0440 ~\pm~ 0.0035 ~\pm~ 0.0038$ \\
$ 37$ & $2.17\cdot10^{-3}$ & $0.905$ & $2.29\cdot10^{-2}$ & $0.0213 ~\pm~ 0.0024 ~\pm~ 0.0037$ \\
 & & & & \\[-1mm] 
$ 37$ & $4.63\cdot10^{-3}$ & $0.365$ & $7.29\cdot10^{-3}$ & $0.0647 ~\pm~ 0.0075 ~\pm~ 0.0053$ \\
$ 37$ & $4.63\cdot10^{-3}$ & $0.455$ & $8.50\cdot10^{-3}$ & $0.0550 ~\pm~ 0.0053 ~\pm~ 0.0039$ \\
$ 37$ & $4.63\cdot10^{-3}$ & $0.545$ & $1.02\cdot10^{-2}$ & $0.0493 ~\pm~ 0.0043 ~\pm~ 0.0036$ \\
$ 37$ & $4.63\cdot10^{-3}$ & $0.635$ & $1.27\cdot10^{-2}$ & $0.0439 ~\pm~ 0.0035 ~\pm~ 0.0029$ \\
$ 37$ & $4.63\cdot10^{-3}$ & $0.725$ & $1.68\cdot10^{-2}$ & $0.0458 ~\pm~ 0.0035 ~\pm~ 0.0027$ \\
$ 37$ & $4.63\cdot10^{-3}$ & $0.815$ & $2.50\cdot10^{-2}$ & $0.0362 ~\pm~ 0.0030 ~\pm~ 0.0017$ \\
$ 37$ & $4.63\cdot10^{-3}$ & $0.905$ & $4.87\cdot10^{-2}$ & $0.0177 ~\pm~ 0.0021 ~\pm~ 0.0038$ \\
 & & & & \\[-1mm] 
\hline
\end{tabular}
}
}
\caption{The semi-inclusive structure function $F_2^{LN(3)}(Q^2,x,x_L)$, for
neutrons with \mbox{$p_T < 0.2~\GeV$},  continued from Table~\ref{tab:f2ln3p2}.}
\label{tab:f2ln3p3}
\end{table}

\begin{table}[p]
\centering
\scalebox{1.25}{
{\tiny
\begin{tabular}{ccccc}
\hline
 & & & & \\[-1mm]
$Q^2 ~[\mathrm{GeV}^2]$  &  $x$  &  $x_L$  &  $\beta$  &  $F_2^{LN(3)}$  \\ 
 & & & & \\[-1mm]
\hline
 & & & & \\[-1mm] 
$ 37$ & $9.87\cdot10^{-3}$ & $0.365$ & $1.55\cdot10^{-2}$ & $0.0469 ~\pm~ 0.0057 ~\pm~ 0.0046$ \\
$ 37$ & $9.87\cdot10^{-3}$ & $0.455$ & $1.81\cdot10^{-2}$ & $0.0446 ~\pm~ 0.0046 ~\pm~ 0.0030$ \\
$ 37$ & $9.87\cdot10^{-3}$ & $0.545$ & $2.17\cdot10^{-2}$ & $0.0390 ~\pm~ 0.0037 ~\pm~ 0.0038$ \\
$ 37$ & $9.87\cdot10^{-3}$ & $0.635$ & $2.70\cdot10^{-2}$ & $0.0372 ~\pm~ 0.0032 ~\pm~ 0.0023$ \\
$ 37$ & $9.87\cdot10^{-3}$ & $0.725$ & $3.59\cdot10^{-2}$ & $0.0335 ~\pm~ 0.0028 ~\pm~ 0.0019$ \\
$ 37$ & $9.87\cdot10^{-3}$ & $0.815$ & $5.34\cdot10^{-2}$ & $0.0326 ~\pm~ 0.0028 ~\pm~ 0.0018$ \\
$ 37$ & $9.87\cdot10^{-3}$ & $0.905$ & $1.04\cdot10^{-1}$ & $0.0138 ~\pm~ 0.0017 ~\pm~ 0.0023$ \\
 & & & & \\[-1mm] 
$ 55$ & $2.17\cdot10^{-3}$ & $0.365$ & $3.42\cdot10^{-3}$ & $0.0962 ~\pm~ 0.0125 ~\pm~ 0.0074$ \\
$ 55$ & $2.17\cdot10^{-3}$ & $0.455$ & $3.99\cdot10^{-3}$ & $0.0621 ~\pm~ 0.0072 ~\pm~ 0.0074$ \\
$ 55$ & $2.17\cdot10^{-3}$ & $0.545$ & $4.77\cdot10^{-3}$ & $0.0727 ~\pm~ 0.0072 ~\pm~ 0.0045$ \\
$ 55$ & $2.17\cdot10^{-3}$ & $0.635$ & $5.95\cdot10^{-3}$ & $0.0622 ~\pm~ 0.0057 ~\pm~ 0.0034$ \\
$ 55$ & $2.17\cdot10^{-3}$ & $0.725$ & $7.90\cdot10^{-3}$ & $0.0540 ~\pm~ 0.0048 ~\pm~ 0.0041$ \\
$ 55$ & $2.17\cdot10^{-3}$ & $0.815$ & $1.17\cdot10^{-2}$ & $0.0486 ~\pm~ 0.0048 ~\pm~ 0.0027$ \\
$ 55$ & $2.17\cdot10^{-3}$ & $0.905$ & $2.29\cdot10^{-2}$ & $0.0204 ~\pm~ 0.0029 ~\pm~ 0.0039$ \\
 & & & & \\[-1mm] 
$ 55$ & $4.63\cdot10^{-3}$ & $0.365$ & $7.29\cdot10^{-3}$ & $0.0695 ~\pm~ 0.0093 ~\pm~ 0.0052$ \\
$ 55$ & $4.63\cdot10^{-3}$ & $0.455$ & $8.50\cdot10^{-3}$ & $0.0614 ~\pm~ 0.0069 ~\pm~ 0.0052$ \\
$ 55$ & $4.63\cdot10^{-3}$ & $0.545$ & $1.02\cdot10^{-2}$ & $0.0448 ~\pm~ 0.0048 ~\pm~ 0.0032$ \\
$ 55$ & $4.63\cdot10^{-3}$ & $0.635$ & $1.27\cdot10^{-2}$ & $0.0482 ~\pm~ 0.0048 ~\pm~ 0.0035$ \\
$ 55$ & $4.63\cdot10^{-3}$ & $0.725$ & $1.68\cdot10^{-2}$ & $0.0498 ~\pm~ 0.0046 ~\pm~ 0.0035$ \\
$ 55$ & $4.63\cdot10^{-3}$ & $0.815$ & $2.50\cdot10^{-2}$ & $0.0415 ~\pm~ 0.0042 ~\pm~ 0.0034$ \\
$ 55$ & $4.63\cdot10^{-3}$ & $0.905$ & $4.87\cdot10^{-2}$ & $0.0164 ~\pm~ 0.0023 ~\pm~ 0.0022$ \\
 & & & & \\[-1mm] 
$ 55$ & $9.87\cdot10^{-3}$ & $0.365$ & $1.55\cdot10^{-2}$ & $0.0592 ~\pm~ 0.0084 ~\pm~ 0.0048$ \\
$ 55$ & $9.87\cdot10^{-3}$ & $0.455$ & $1.81\cdot10^{-2}$ & $0.0429 ~\pm~ 0.0053 ~\pm~ 0.0027$ \\
$ 55$ & $9.87\cdot10^{-3}$ & $0.545$ & $2.17\cdot10^{-2}$ & $0.0409 ~\pm~ 0.0046 ~\pm~ 0.0030$ \\
$ 55$ & $9.87\cdot10^{-3}$ & $0.635$ & $2.70\cdot10^{-2}$ & $0.0372 ~\pm~ 0.0038 ~\pm~ 0.0027$ \\
$ 55$ & $9.87\cdot10^{-3}$ & $0.725$ & $3.59\cdot10^{-2}$ & $0.0338 ~\pm~ 0.0034 ~\pm~ 0.0022$ \\
$ 55$ & $9.87\cdot10^{-3}$ & $0.815$ & $5.34\cdot10^{-2}$ & $0.0304 ~\pm~ 0.0032 ~\pm~ 0.0018$ \\
$ 55$ & $9.87\cdot10^{-3}$ & $0.905$ & $1.04\cdot10^{-1}$ & $0.0148 ~\pm~ 0.0022 ~\pm~ 0.0037$ \\
 & & & & \\[-1mm] 
$ 55$ & $2.10\cdot10^{-2}$ & $0.365$ & $3.31\cdot10^{-2}$ & $0.0515 ~\pm~ 0.0089 ~\pm~ 0.0046$ \\
$ 55$ & $2.10\cdot10^{-2}$ & $0.455$ & $3.86\cdot10^{-2}$ & $0.0347 ~\pm~ 0.0051 ~\pm~ 0.0021$ \\
$ 55$ & $2.10\cdot10^{-2}$ & $0.545$ & $4.62\cdot10^{-2}$ & $0.0286 ~\pm~ 0.0038 ~\pm~ 0.0038$ \\
$ 55$ & $2.10\cdot10^{-2}$ & $0.635$ & $5.76\cdot10^{-2}$ & $0.0334 ~\pm~ 0.0040 ~\pm~ 0.0018$ \\
$ 55$ & $2.10\cdot10^{-2}$ & $0.725$ & $7.65\cdot10^{-2}$ & $0.0303 ~\pm~ 0.0036 ~\pm~ 0.0020$ \\
$ 55$ & $2.10\cdot10^{-2}$ & $0.815$ & $1.14\cdot10^{-1}$ & $0.0266 ~\pm~ 0.0033 ~\pm~ 0.0023$ \\
$ 55$ & $2.10\cdot10^{-2}$ & $0.905$ & $2.21\cdot10^{-1}$ & $0.0098 ~\pm~ 0.0018 ~\pm~ 0.0018$ \\
 & & & & \\[-1mm] 
$ 82$ & $4.63\cdot10^{-3}$ & $0.365$ & $7.29\cdot10^{-3}$ & $0.0588 ~\pm~ 0.0099 ~\pm~ 0.0045$ \\
$ 82$ & $4.63\cdot10^{-3}$ & $0.455$ & $8.50\cdot10^{-3}$ & $0.0537 ~\pm~ 0.0080 ~\pm~ 0.0036$ \\
$ 82$ & $4.63\cdot10^{-3}$ & $0.545$ & $1.02\cdot10^{-2}$ & $0.0465 ~\pm~ 0.0061 ~\pm~ 0.0036$ \\
$ 82$ & $4.63\cdot10^{-3}$ & $0.635$ & $1.27\cdot10^{-2}$ & $0.0520 ~\pm~ 0.0063 ~\pm~ 0.0052$ \\
$ 82$ & $4.63\cdot10^{-3}$ & $0.725$ & $1.68\cdot10^{-2}$ & $0.0480 ~\pm~ 0.0056 ~\pm~ 0.0033$ \\
$ 82$ & $4.63\cdot10^{-3}$ & $0.815$ & $2.50\cdot10^{-2}$ & $0.0417 ~\pm~ 0.0053 ~\pm~ 0.0040$ \\
$ 82$ & $4.63\cdot10^{-3}$ & $0.905$ & $4.87\cdot10^{-2}$ & $0.0180 ~\pm~ 0.0032 ~\pm~ 0.0028$ \\
 & & & & \\[-1mm] 
$ 82$ & $9.87\cdot10^{-3}$ & $0.365$ & $1.55\cdot10^{-2}$ & $0.0484 ~\pm~ 0.0085 ~\pm~ 0.0041$ \\
$ 82$ & $9.87\cdot10^{-3}$ & $0.455$ & $1.81\cdot10^{-2}$ & $0.0514 ~\pm~ 0.0077 ~\pm~ 0.0036$ \\
$ 82$ & $9.87\cdot10^{-3}$ & $0.545$ & $2.17\cdot10^{-2}$ & $0.0366 ~\pm~ 0.0052 ~\pm~ 0.0025$ \\
$ 82$ & $9.87\cdot10^{-3}$ & $0.635$ & $2.70\cdot10^{-2}$ & $0.0507 ~\pm~ 0.0064 ~\pm~ 0.0038$ \\
$ 82$ & $9.87\cdot10^{-3}$ & $0.725$ & $3.59\cdot10^{-2}$ & $0.0361 ~\pm~ 0.0044 ~\pm~ 0.0027$ \\
$ 82$ & $9.87\cdot10^{-3}$ & $0.815$ & $5.34\cdot10^{-2}$ & $0.0311 ~\pm~ 0.0041 ~\pm~ 0.0029$ \\
$ 82$ & $9.87\cdot10^{-3}$ & $0.905$ & $1.04\cdot10^{-1}$ & $0.0170 ~\pm~ 0.0030 ~\pm~ 0.0027$ \\
 & & & & \\[-1mm] 
$ 82$ & $2.10\cdot10^{-2}$ & $0.365$ & $3.31\cdot10^{-2}$ & $0.0477 ~\pm~ 0.0093 ~\pm~ 0.0040$ \\
$ 82$ & $2.10\cdot10^{-2}$ & $0.455$ & $3.86\cdot10^{-2}$ & $0.0319 ~\pm~ 0.0052 ~\pm~ 0.0025$ \\
$ 82$ & $2.10\cdot10^{-2}$ & $0.545$ & $4.62\cdot10^{-2}$ & $0.0366 ~\pm~ 0.0053 ~\pm~ 0.0024$ \\
$ 82$ & $2.10\cdot10^{-2}$ & $0.635$ & $5.76\cdot10^{-2}$ & $0.0372 ~\pm~ 0.0050 ~\pm~ 0.0030$ \\
$ 82$ & $2.10\cdot10^{-2}$ & $0.725$ & $7.65\cdot10^{-2}$ & $0.0443 ~\pm~ 0.0057 ~\pm~ 0.0028$ \\
$ 82$ & $2.10\cdot10^{-2}$ & $0.815$ & $1.14\cdot10^{-1}$ & $0.0241 ~\pm~ 0.0035 ~\pm~ 0.0020$ \\
$ 82$ & $2.10\cdot10^{-2}$ & $0.905$ & $2.21\cdot10^{-1}$ & $0.0132 ~\pm~ 0.0026 ~\pm~ 0.0027$ \\
 & & & & \\[-1mm] 
\hline
\end{tabular}
}
}
\caption{The semi-inclusive structure function $F_2^{LN(3)}(Q^2,x,x_L)$, for
neutrons with \mbox{$p_T<0.2~\GeV$},  continued from Table~\ref{tab:f2ln3p3}.}
\label{tab:f2ln3p4}
\end{table}

\newpage

\end{document}